\documentclass[]{elsart}               

\usepackage{amsmath}
\usepackage{amssymb}
\usepackage{graphicx}
\usepackage{subfigure}
\usepackage{cite}
\usepackage{alltt}
\usepackage{bm}

\begin{document}
\begin{frontmatter}

\title{{\tt THERMINATOR 2}: {\tt THERM}al heavy {\tt I}o{\tt N} gener{\tt ATOR} 2\thanksref{grant}} 

\thanks[grant]{Supported in part by the Polish Ministry of Science and Higher Education, grant  No. N N202 263438 and Foundation for Polish Science.}

\author[ifj]{Miko\l{}aj Chojnacki},
\ead{Mikolaj.Chojnacki@ifj.edu.pl}
\author[ohio,pw]{Adam Kisiel},
\ead{kisiel@if.pw.edu.pl}
\author[ifj,as]{Wojciech Florkowski},
\ead{Wojciech.Florkowski@ifj.edu.pl}
\author[ifj,as]{Wojciech Broniowski},
\ead{Wojciech.Broniowski@ifj.edu.pl}

\address[ifj]{The H.~Niewodnicza\'nski Institute of Nuclear Physics, Polish Academy of Sciences, PL-31342 Cracow, Poland}
\address[ohio]{Physics Department, CERN, CH-1211 Geneve 23, Switzerland}
\address[pw]{Faculty of Physics, Warsaw University of Technology, PL-00661~Warsaw, Poland}
\address[as]{Institute of Physics, Jan Kochanowski University, PL-25406~Kielce, Poland} 

\begin{abstract}
We present an extended version of {\tt THERMINATOR}, a Monte Carlo event generator dedicated to studies of the statistical production of particles in relativistic heavy-ion collisions. The increased functionality of the code contains the following features: The input of any shape of the freeze-out hypersurface and the expansion velocity field, including the 3+1 dimensional profiles, in particular those generated externally with various hydrodynamic codes. The hypersufraces may have variable thermal parameters, which allows for studies departing significantly from the mid-rapidity region, where the baryon chemical potential becomes large. We include a library of standard sets of hypersurfaces and velocity profiles describing the RHIC Au+Au data at $\sqrt{s_{NN}}=200$~GeV for various centralities, as well as those anticipated for the LHC Pb+Pb collisions at $\sqrt{s_{NN}}=5.5$~TeV. A separate code, {\tt FEMTO-THERMINATOR}, is provided to carry out the analysis of femtoscopic correlations which are an important source of information concerning the size and expansion of the system. We also include several useful 
scripts that carry out auxiliary tasks, such as obtaining an estimate of the number of elastic collisions after the freeze-out, counting of particles flowing back into the fireball and violating causality (typically very few), or visualizing various results: the particle $p_T$-spectra, the elliptic flow coefficients, and the HBT correlation radii. We also investigate the problem of the back-flow 
of particles into the hydrodynamic region, as well as estimate the elastic rescattering in terms of trajectory crossings. 
The package is written in {\tt C++} and uses the CERN {\tt ROOT} environment. 
\end{abstract}

\date{Jan. 30, 2011}


\end{frontmatter}

\maketitle

\newpage

\noindent{\bf Program summary}

\noindent
{\sl Title of the program:}\-  {\tt THERMINATOR 2}\\
{\sl Catalog identifier:}\- \\
{\sl Program web page:}\\ \- http://therminator2.ifj.edu.pl/ \\
{\sl Licensing provisions:}\- none \\
{\sl Computer:} \- any computer with a {\tt C++} compiler and the CERN {\tt ROOT} environment, ver.~5.26 or later, 
 tested with Intel Core2 Duo CPU E8400 @ 3~GHz, 4~GB RAM\\
{\sl Operating system under which the program has been tested:} \-\\
{Linux} Ubuntu~10.10 x64 ({\tt gcc} 4.4.5) {\tt ROOT}~ver.~5.26\\
{\sl Programming language used:} \- {\tt C++} with the CERN {\tt ROOT} libraries, BASH shell\\
{\sl No. of lines in distributed program, including test data:} \-\\
13000 code, 300 configuration, 400000 input files (hypersurface library)\\
{\sl No. of bytes in distributed program, including test data:} \- 2.7~MB\\
{\sl Distribution format:} \- tar.gz\\
{\sl Nature of physical problem:} \-  particle production via statistical hadronization in relativistic heavy-ion collisions\\
{\sl Method of solution:} \-  Monte-Carlo simulation, analyzed with {\tt ROOT}\\
{\sl Restrictions concerning the complexity of the problem:} \- none\\
{\sl Memory required to execute with typical data:} \-\\
--- 30~MB {\tt therm2\_events},\\
--- 150~MB {\tt therm2\_femto}\\
{\sl Typical running time:} \- default configuration at 3~GHz\\
--- primordial multiplicities 70~min (calculated only once per case),\\
--- 8~min/500~events,\\
--- 10~min -- draw all figures,\\
--- 25~min/one $k_T$ bin in the HBT analysis with 5000 events\\
{\sl Typical data file size:} \- default configuration\\
--- 45~MB/500~events\\
--- 35~MB/correlation file (one $k_T$ bin)\\
--- 45~kB/fit file (projections and fits)\\

\section{Introduction \label{sec:intro}}

We introduce an updated and largely extended version of {\tt THERMINATOR} \cite{Kisiel:2005hn}, the {\tt THERM}al heavy {\tt I}o{\tt N} gener{\tt ATOR}, created to carry out the statistical hadronization in relativistic heavy-ion collisions. Numerous successful analyses have been performed with the help of our code over the last few years \cite{Broniowski:2005ae,Pratt:2005bt,Andronic:2005yp,Kisiel:2006is,Chojnacki:2006tv,Florkowski:2006mb,Biedron:2006vf,
Utyuzh:2006nw,Begun:2006uu,Utyuzh:2007ct,Brown:2007raa,Baeuchle:2007et,Vertesi:2007ki,Danielewicz:2007jn,%
Begun:2007du,Tomasik:2007gs,Chojnacki:2007rq,Bozek:2007qt,Afanasiev:2007kk,Akiba:2008zz,Broniowski:2008vp,%
Florkowski:2008cs,Lacey:2008kd,Chung:2008fu,Kisiel:2008ws,:2008fq,Tomasik:2008fq,Bozek:2008zw,:2008gk,Broniowski:2008qk,%
Broniowski:2008ee,Florkowski:2009vr,Chajecki:2009zg,Bozek:2009ty,Luo:2009sx,Bozek:2009zz,NoronhaHostler:2009tz,%
Broniowski:2009fm,Florkowski:2009rt,Bozek:2009mz,Hauer:2009nu,Ryblewski:2009hm,Broniowski:2009te,%
Videbaek:2009zy,Bozek:2009dt,Kisiel:2009iw,Kisiel:2009eh,Kisiel:2009kn,Florkowski:2009wb,Heinz:2009xj,Bozek:2009pu,Lokhtin:2009hs,NoronhaHostler:2010yc,Luo:2010by,Bozek:2010bi,Aggarwal:2010wy,Hauer:2010sv,NoronhaHostler:2010hy,Bozek:2010aj,Bilandzic:2010jr,Lizhu:2010zh,Bozek:2010vz}, 
contributing to better understanding of the involved evolution of the hot and dense system created in ultra-relativistic heavy-ion collision.

{\tt THERMINATOR 2} is a Monte Carlo generator written in {\tt C++} and using the standard CERN {\tt ROOT} \cite{root} environment. That way, apart from model applications, the code can be easily adapted for purposes directly linked to experimental data analysis, detector modeling, or estimates for the  heavy-ion experiments at RHIC,  LHC, SPS, FAIR, or NICA.

Originally, {\tt THERMINATOR} was designed to perform the hadronic freeze-out on simple boost-invariant hypersurfaces, such as the Cracow single-freeze-out model~\cite{Florkowski:2001fp,Broniowski:2001we} and the Blast-Wave model~\cite{Danielewicz:1992mi,Schnedermann:1993ws,Retiere:2003kf}. Meanwhile, the code has evolved into a versatile tool, where the freeze-out profile and the expansion velocity field of any shape can be implemented, allowing application to all approaches based on statistical hadronization on a specified hypersurface. The present functionality of the code includes the following new features:

\begin{enumerate}

\item An implementation of any shape of the freeze-out hypersurface and the expansion velocity field is possible now, including the 2+1 and 3+1 dimensional profiles, in particular those generated externally with various codes for perfect hydrodynamics \cite{Teaney:2001av,Hirano:2002ds,Kolb:2003dz,Huovinen:2003fa,Shuryak:2004cy,Eskola:2005ue,Hama:2005dz,%
Hirano:2005xf,Huovinen:2006jp,Hirano:2007xd,Nonaka:2006yn,Broniowski:2008vp,Huovinen:2009yb}.  

\item The hypersufraces may have space-time dependent thermal parameters, which allows for studies of non-boost-invariant systems. In particular, one may depart significantly from the mid-rapidity region to the fragmentation regions where the baryon chemical potential becomes large~\cite{Biedron:2006vf,Videbaek:2009zy}.

\item The package includes a library of ``standard'' sets of hypersurfaces and velocity profiles, which describe the Au+Au data \cite{Broniowski:2008vp} at the highest RHIC energy $\sqrt{s_{NN}}=200$~GeV for various centralities. We also provide the hypersurfaces and velocity profiles anticipated \cite{Florkowski:2008cs} for the LHC Pb+Pb collisions at $\sqrt{s_{NN}}=5.5$~TeV, which prior to studies with the real data can be used for the modeling of detectors.

\item A separate code, the {\tt FEMTO-THERMINATOR}, is provided to carry out the analysis of the femtoscopic correlations. These correlations are an important source of complementary information concerning the size and expansion of the system, which should be reproduced in a realistic description of the data \cite{Kisiel:2006is,Broniowski:2008vp,Kisiel:2008ws}.

\end{enumerate}

We also provide several scripts that carry out auxiliary tasks:

\begin{enumerate}

\item Counting the number of particles flowing back into the fireball, thus violating causality (for realistic freeze-out 
profiles this number is negligible, below 1\%).

\item An estimate of the number of elastic collisions (trajectory crossings) after the freeze-out. For realistic cases this number is small ($\sim 1.5$ for pions scattering off other pions), supporting the single-freeze-out approximation.

\item Visualization of the results: the particle $p_T$-spectra, the elliptic flow coefficient $v_2$, or the HBT correlation radii.

\end{enumerate}

We have also made an effort to make the package more user-friendly, providing numerous examples of usage in typical situations.

Other codes implementing the thermal description of hadronization have also been developed recently, to mention {\tt SHARE} \cite{Torrieri:2004zz,Torrieri:2006xi} and {\tt THERMUS} \cite{Wheaton:2004qb}, computing the particle abundances, {\tt DRAGON} \cite{Tomasik:2008fq}, a Monte Carlo generator of particle production from a fragmented fireball, or {\tt HYDJET++} \cite{Lokhtin:2008xi,Lokhtin:2009be,Lokhtin:2009hs}, a heavy-ion event generator with hydrodynamics and jets. A code similar to {\tt THERMINATOR}, {\tt Fast Hadron Freeze-out Generator},  has been presented in Refs. \cite{Amelin:2006qe,Amelin:2007ic}.

\section{Statistical hadronization \label{sec:sh}}

For the detailed description of the physics behind {\tt THERMINATOR 2} the reader is referred to the original paper \cite{Kisiel:2005hn}. Here we only wish to stress that the statistical approach to heavy-ion collisions has proved to be very useful for a very broad range of observables. With few physical input parameters, such as the temperature, chemical potentials, size, and the velocity of the collective flow, the models describe the observed particle abundances \cite{Koch:1985hk,Cleymans:1992zc,Sollfrank:1993wn,Schnedermann:1993ws,Braun-Munzinger:1994xr,Braun-Munzinger:1995bp,Csorgo:1995bi,Cleymans:1996cd,Rafelski:1996hf,Rafelski:1997ab,Becattini:1997uf,Yen:1998pa,Cleymans:1998fq,Gazdzicki:1998vd,Gazdzicki:1999ej,Braun-Munzinger:1999qy,Cleymans:1999st,Becattini:2000jw,Braun-Munzinger:2001ip,Florkowski:2001fp,Broniowski:2002nf,Retiere:2003kf}, the transverse-momentum spectra \cite{Broniowski:2001we}, the balance functions \cite{Florkowski:2004em,Bozek:2003qi}, the elliptic flow coefficient \cite{Broniowski:2002wp,Florkowski:2004du}, and femtoscopic observables \cite{Kisiel:2006is,Broniowski:2008vp,Kisiel:2008ws}. A uniform description of the spectra, the elliptic flow coefficient $v_2$, and the HBT data achieved in \cite{Broniowski:2008vp} explained for the first time the so-called ``RHIC HBT puzzle''.

The key element of the success behind the statistical approach is the inclusion of the full list \cite{Amsler:2008zzb} of hadronic resonances, whose number grows rapidly according to the Hagedorn hypothesis \cite{Hagedorn:1965st,Hagedorn:1968ua,Hagedorn:1994sc,Broniowski:2000bj,Broniowski:2000hd,Broniowski:2004yh}. At the rather high temperature of the freeze-out, $\sim 140-165$~MeV, the resonances contribute very significantly to the observables. {\tt THERMINATOR 2} uses the particle data tables in the universal form originally included in the {\tt SHARE} \cite{Torrieri:2004zz} package, and incorporates the four-**** and three-*** resonances.

All particles are created at the freeze-out hypersurface according to the Cooper-Frye formula \cite{Cooper:1974mv}. The two- and three-body decays of resonances, proceeding in cascades, produce the stable particles observed in the detectors. {\tt THERMINATOR} offers the {full space-time information on positions and velocities of the produced particles}. As a Monte Carlo event generator written in {\tt C++} in the CERN {\tt ROOT} \cite{root} environment,  {\tt THERMINATOR} can be straightforwardly interfaced to the standard software routinely used in the data analysis of relativistic heavy-ion collisions at the LHC, RHIC, SPS, FAIR or NICA. The inclusion of experimental acceptance and kinematic cuts poses no difficulty in a Monte Carlo generator.

\subsection{Cooper-Frye formalism \label{sec:cf-form}}
%
Our starting point for the analysis of different models is the famous Cooper-Frye formula~\cite{Cooper:1974mv}. It expresses the number of hadrons being produced on the freeze-out hypersurface $\Sigma^\mu$ by the following integral
\begin{equation}
N = (2s+1)\int \frac{d^3p}{(2\pi)^3 E_p} \int d\Sigma_\mu(x) p^\mu f(x,p),
\label{cp1}
\end{equation}
where
\begin{equation}
  f(p\cdot u) = \left\{ \exp\left[\frac{p_\mu u^\mu - \left(  B \mu_B + I_3 \mu_{I_3} + S \mu_S + C \mu_C \right)}{T}\right] \pm 1 \right\}^{-1}
  \label{eqn_PdotU0}
\end{equation}
is the phase-space distribution function of the particles (the stable ones and resonances), and $d\Sigma^\mu$ is the three-dimensional element of the freeze-out hypersurface. The latter may be calculated with the help of the formula known from the differential geometry \cite{Misner:1974qy}
\begin{equation}
d\Sigma_\mu = \varepsilon_{\mu \alpha \beta \gamma}
\frac{\partial x^\alpha}{\partial\alpha} \frac{\partial x^\beta}{\partial\beta} \frac{\partial x^\gamma}{\partial\gamma }
d\alpha d\beta d\gamma,
\label{d3Sigma}
\end{equation}
where $\varepsilon_{\mu \alpha \beta \gamma}$ is the Levi-Civita tensor ($\varepsilon_{0123} = +1$) and the variables $\alpha$, $\beta$, and $\gamma$ are used to parametrize the three-dimensional freeze-out hypersurface submerged in the four-dimensional Minkowski space. The quantity $d\Sigma^\mu$ may be interpreted as the four-vector normal to the hypersurface with the norm equal to the ``volume of the hypersurface element''. Explicitly,
\begin{eqnarray}
d\Sigma_0 =
\left|
\begin{array}{ccc}
\frac{\partial x}{\partial \alpha}  &\frac{\partial x}{\partial \beta} &\frac{\partial x}{\partial \gamma} \\
\frac{\partial y}{\partial \alpha}  &\frac{\partial y}{\partial \beta} &\frac{\partial y}{\partial \gamma} \\
\frac{\partial z}{\partial \alpha}  &\frac{\partial z}{\partial \beta} &\frac{\partial z}{\partial \gamma}
\end{array}
\right| d\alpha d\beta d\gamma,
\label{dSigmadet2}
\end{eqnarray}
while the remaining components are obtained via cyclic permutations of $t$, $z$, $y$, and $z$.
In the following subsections we shall explicitly provide $d\Sigma_\mu$  in the form appropriate for various considered cases. The formulas will be directly used to generate the hadrons with the Monte-Carlo method. 
\begin{figure}[t]
\begin{center}
\includegraphics[width=0.4\textwidth]{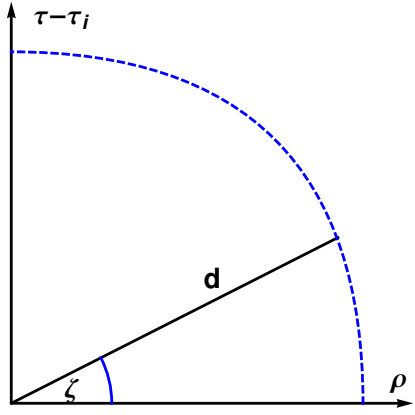}
\end{center}
\caption{The system of coordinates used to parametrize the boost-invariant freeze-out hypersurfaces obtained from the 2+1 hydrodynamic codes. The dashed line shows the freeze-out points with the fixed values of the azimuthal angle $\phi$. }
\label{fig:2d}
\end{figure}

The quantities appearing in the expressions below have the following physical interpretation:
\begin{itemize}
  \item thermodynamic quantities on the freeze-out hypersurface:
  \begin{itemize}
    \medskip
    \item $T$ - temperature,
    \item $\mu_B$ - baryon chemical potential,
    \item $\mu_{I_3}$ - isospin chemical potential,
    \item $\mu_S$ - strange chemical potential,
    \item $\mu_C$ - charmed chemical potential, 
    \medskip
  \end{itemize}
  \item space-time coordinates:
  \begin{itemize}
    \medskip
   \item $\tau = \sqrt{t^2-z^2}$ - longitudinal proper time,
   \item $\rho = \sqrt{x^2+y^2}$ - distance in the transverse plane,
   \item $\phi$ - azimuthal angle,
   \item $Y_s = \frac{1}{2} \ln ((t+z)/(t-z))$ - space-time rapidity,
    \medskip
  \end{itemize}
  \item coordinates and parameters used to specify the freeze-out hypersurface:
  \begin{itemize}
      \medskip
    \item $\tau_i$ - the initial proper time for hydrodynamics,
    \item $\tau_f$ - the final proper time,
    \item $\zeta$ - angle in $\rho-\tau$ plane, see Fig.~\ref{fig:2d}
    \item $\theta$ - angle between the $Y_s$-axis and the direction determined by the space time point ($\tau_i$,0,0,0) and the point on the freeze-out hypersurface, see Fig.~\ref{fig:3d},
    \item $d = d(\zeta,\phi,\theta)$ - distance from the space-time point ($\tau_i$,0,0,0) to the point on the freeze-out hypersurface, see Fig.~\ref{fig:3d},
     \medskip
  \end{itemize}
  \item flow characteristics of the fluid element:
  \begin{itemize}
      \medskip
    \item $u_x$, $u_y$ - transverse components of the four-velocity,
    \item $v_T = \sqrt{v_x^2+v_y^2}$ - magnitude of the transverse velocity,
    \item $\phi_f$ - azimuthal angle of the transverse velocity,
    \item $Y_f$ - rapidity of the fluid element,
     \medskip
  \end{itemize}
  \item properties of a particle generated on the freeze-out hypersurface:
  \begin{itemize}
      \medskip
    \item $s$ - spin,
    \item $m$ - mass,
    \item $p_T$ - transverse momentum,
    \item $m_T = \sqrt{m^2 + p_T^2}$ - transverse mass,
     \item $\phi_p$ - azimuthal angle of the transverse momentum,
    \item $Y$ - rapidity,
  \end{itemize}
\end{itemize}
In the case of freeze-out hypersurfaces obtained in the framework of the perfect-fluid hydrodynamics the distribution function in (\ref{cp1}) is the Fermi-Dirac or Bose-Einstein distribution and the number of hadrons produced is given by the formula
\begin{eqnarray}
\hspace{-0.75cm}  N = (2 s + 1) \int  \frac{dY\, p_T\, dp_T\, d\phi_p}{(2 \pi)^3}\, \int \, 
\frac{d\Sigma_\mu p^\mu\,}{\exp(-\frac{B \mu_B + I_3 \mu_{I_3} + S \mu_S + C \mu_C}{T})\exp(\frac{p_\mu u^\mu}{T}) \pm 1}.
\nonumber \\
\end{eqnarray}
%
\subsection{Freeze-out in 3+1-dimensional perfect-fluid hydrodynamics \label{sec:3plus1}}
%
\begin{figure}[b]
\begin{center}
\includegraphics[width=7.5cm]{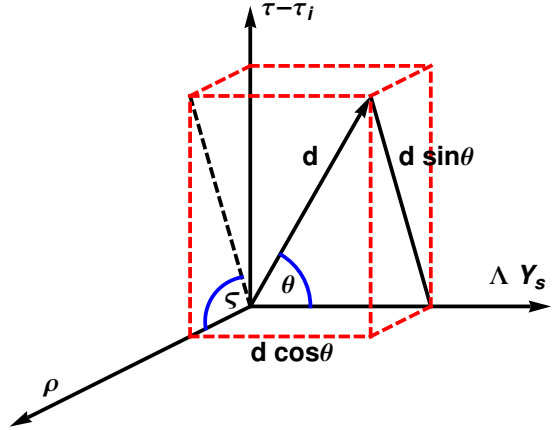}
\end{center}
\caption{The system of coordinates used to parametrize the freeze-out hypersurfaces obtained from the 3+1 hydrodynamic codes, see Eqs.~(\ref{3d-par1}) and (\ref{3d-par2}). }
\label{fig:3d}
\end{figure}
First, we consider the most general case where both the freeze-out hypersurface and the hydrodynamic flow at freeze-out are obtained from the 3+1 perfect-fluid hydrodynamic code, with no symmetries assumed. In this situation the freeze-out hypersurface is parametrized in the following way \cite{Bozek:2009ty},
\begin{eqnarray}
  t &=& \left(\tau_i + d(\zeta,\phi,\theta) \sin\theta \sin\zeta \right)
         \cosh\frac{d(\zeta,\phi,\theta) \cos\theta}{\Lambda}, \nonumber \\
  x &=& d(\zeta,\phi,\theta) \sin\theta \cos\zeta \cos\phi, \nonumber \\
  y &=& d(\zeta,\phi,\theta) \sin\theta \cos\zeta \sin\phi, \nonumber \\
  z &=& \left(\tau_i + d(\zeta,\phi,\theta) \sin\theta \sin\zeta \right)
          \sinh\frac{d(\zeta,\phi,\theta) \cos\theta}{\Lambda},
\label{3d-par1}
\end{eqnarray}
where $d$ is a function of the parameters $\phi, \zeta$, and $\theta$, which is obtained numerically from the hydrodynamic code, see Fig.~\ref{fig:3d}. The parameter $\Lambda$ is a scale used to change the dimensionless space-time rapidity $Y_s$ into a dimensional quantity (for example one may choose  $\Lambda = \rho_{\rm max}/Y_s^{\rm max}$), namely
\begin{eqnarray}
Y_s &=& Y_s(\zeta,\phi,\theta) = \frac{d(\zeta,\phi,\theta) \cos\theta}{\Lambda},
\nonumber \\
 \tau &=& \tau(\zeta,\phi,\theta) = \tau_i + d(\zeta,\phi,\theta) \sin\theta \sin\zeta.
\label{3d-par2}
\end{eqnarray}
In addition, we express the particle four-momentum and the fluid four-velocity in terms of particle and fluid rapidities, respectively, which leads to the expression
\begin{equation}
p_\mu u^\mu = \sqrt{1 + u_x^2 + u_y^2}\, \left[ m_T \cosh (Y_f - Y) - p_T ( u_x \cos\phi_p + u_y \sin\phi_p ) \right].
\label{3d-pu}
\end{equation}
The calculation of the integration measure in the Cooper-Frye gives
\begin{eqnarray}
  d\Sigma_\mu p^\mu 	&=& d^2 \sin\theta\,
\Biggl\{ \Biggr. \frac{\left(\tau_i + d \sin\zeta\right)}{\Lambda}  \Biggl[ \Biggr.  \frac{\partial d}{\partial\zeta} \cos\zeta \Biggl( \Biggr.  - m_T \cos\zeta \cosh\left(Y - \frac{d \cos\theta}{\Lambda}\right)  \nonumber\\
			& & \qquad
				+ p_T \sin\zeta \cos(\phi - \phi_p) \Biggl.  \Biggr)
+ \left(d \sin\theta - \frac{\partial d}{\partial\theta} \cos\theta\right)
\nonumber\\
			& & \times \left(
				  m_T \sin\zeta \cosh\left(Y - \frac{d \cos\theta}{\Lambda}\right)\quad
				+ p_T \cos\zeta \cos(\phi - \phi_p)
			     \right) \nonumber\\
			& & \quad + \frac{\partial d}{\partial\phi}\,\, p_T\, \sin(\phi - \phi_p) \Biggl. \Biggr] \nonumber\\
			&&  + \left(d \cos\theta + \frac{\partial d}{\partial\theta} \sin\theta \right) \sin\theta \sinh\left(Y - \frac{d \cos\theta}{\Lambda}\right) \Biggl.  \Biggr\} .
\label{3d-Sp1}
\end{eqnarray}
Using (\ref{3d-par2}) in (\ref{3d-Sp1}) one obtains a more compact form
\begin{eqnarray}
  d\Sigma_\mu p^\mu 	&=& \frac{d \sin\theta}{\Lambda}\, d\, \tau \Biggl[ \Biggr.  \frac{\partial d}{\partial\zeta} \cos\zeta \left(
				- m_T \cos\zeta \cosh\left(Y - Y_s\right) \right.
				 \nonumber\\
			&& \left. + p_T \sin\zeta \cos\left(\phi - \phi_p\right)
			    \right) + \cos\zeta \sin\theta \left(d \sin\theta - \frac{\partial d}{\partial\theta} \cos\theta\right) \nonumber \\
&& \times \left(
				  m_T \sin\zeta \cosh\left(Y -Y_s\right)
				+ p_T \cos\zeta \cos\left(\phi - \phi_p\right)
			    \right) \nonumber\\
			&& \quad + \frac{\partial d}{\partial\phi}\,\, p_T\, \sin(\phi - \phi_p) \nonumber\\
			&& \quad + \frac{\Lambda}{\tau} \left(d \cos\theta + \frac{\partial d}{\partial\theta} \sin\theta \right) \sin\theta \sinh\left(Y - Y_s\right)  \Biggl. \Biggr].
\label{3d-Sp2}
\end{eqnarray}

\subsection{Freeze-out in 2+1-dimensional boost-invariant hydrodynamics \label{sec:2plus1}}
%
For boost-invariant systems, considered in approximate treatment of the heavy-ion collisions, the fluid rapidity is equal to space-time rapidity, $Y_f = Y_s$, and the function $d$ depends only on $\phi$ and $\zeta$. In this case we find \cite{Chojnacki:2007rq}
\begin{eqnarray}
  p_\mu u^\mu &=& \frac{1}{\sqrt{1-v_T^2}} (m_T \cosh(Y - Y_s) - v_T\, p_T \cos(\phi - \phi_p))
\label{2d-pu}
\end{eqnarray}
and
\begin{eqnarray}
  d\Sigma_\mu p^\mu 	&=& d\, \tau \Biggl[ \Biggr.  \frac{\partial d}{\partial\zeta} \cos\zeta \left(
				- m_T \cos\zeta \cosh\left(Y - Y_s\right)
				+ p_T \sin\zeta \cos\left(\phi - \phi_p\right)
			    \right) \nonumber\\
			&& \quad + d \cos\zeta \left(
				  m_T \sin\zeta \cosh\left(Y -Y_s\right)
				+ p_T \cos\zeta \cos\left(\phi - \phi_p\right)
			    \right) \nonumber\\
			&& \quad + \frac{\partial d}{\partial\phi}\,\, p_T\, \sin(\phi - \phi_p)  \Biggl. \Biggr].
\label{2d-Sp}
\end{eqnarray}
%
\subsection{Blast-wave models  \label{sec:model-a}}

%
\begin{figure}[t]
\begin{center}
\includegraphics[width=0.5\textwidth]{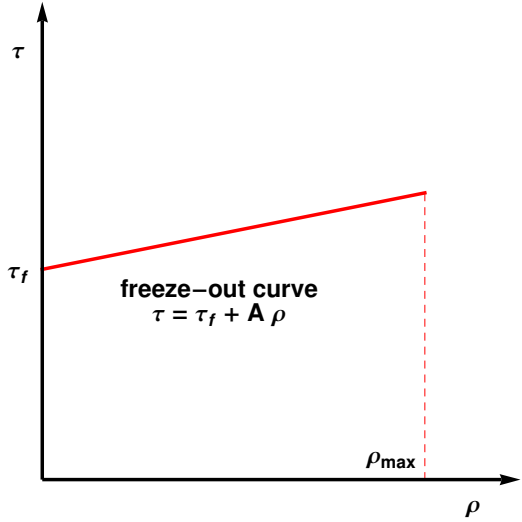}
\end{center}
\caption{The freeze-out curves considered in the modified Blast-Wave model \cite{Kisiel:2006is}. }
\label{fig:model-a}
\end{figure}
In this Section we discuss one of the most popular parameterizations of the freeze-out hypersurface --- the Blast-Wave model. In its standard form, the model is boost-invariant and cylindrically symmetric. Moreover, the assumption is made that the freeze-out happens at a constant value of the proper time
\begin{equation}
\tau(\zeta) = \tau_{f} = \hbox{const}.
\label{model-a1}
\end{equation}
In order to get a broader applicability, we generalize this condition to the formula
\begin{equation}
\tau(\rho) = \tau_f  + A \rho,
\label{model-a2}
\end{equation}
where $A$ is a constant that describes the slope of the freeze-out curve in the Minkowski space, see Fig.~\ref{fig:model-a}. With $A>0$ ($A<0$) we may consider the freeze-out scenarios where the outer parts of the system freeze-out later (earlier). Of course, with $A=0$ we reproduce the standard Blast-Wave parametrization.
In this case we use
\begin{eqnarray}
p_\mu u^\mu &=& \frac{1}{\sqrt{1-v_T^2}} (m_T \cosh(Y_s - Y) - v_T\, p_T \cos(\phi - \phi_p))
\label{A-pu}
\end{eqnarray}
and
\begin{eqnarray}
  d\Sigma_\mu p^\mu 	&=& (\tau_f + A\, \rho)\, \rho\, \left(m_T \cosh(Y_s - Y) - A\, p_T \cos(\phi - \phi_p) \right).
\label{A-Sp}
\end{eqnarray}
The freeze-out conditions defined by Eq. (\ref{model-a2}) with different values of $A$ were studied in \cite{ Kisiel:2006is}.

\subsection{Cracow model  \label{sec:cracow}}

\begin{figure}[t]
\begin{center}
\includegraphics[angle=0,width=0.5\textwidth]{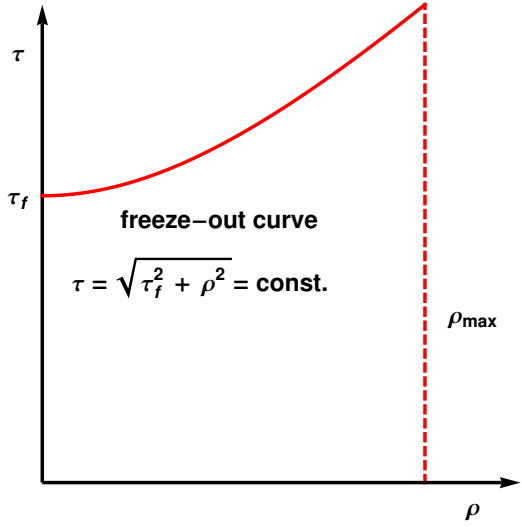}
\end{center}
\caption{The freeze-out curve assumed in the Cracow model.}
\label{fig:hubble}
\end{figure}

%
Another boost-invariant and cylindrically symmetric freeze-out model \cite{Broniowski:2001we} assumes that the hypersurface is defined by the condition, 
\begin{equation}
x^\mu x_\mu = t^2 - x^2 - y^2 - z^2 =
\tau^2 -\rho^2 = \tau_f^2 = \hbox{const.},
\label{hubtau}
\end{equation}
see Fig.~\ref{fig:hubble}, and the fluid four-velocity is proportional to the space-time position
\begin{equation}
u^\mu = \gamma ( 1, {\bf v}) = {x^\mu \over \tau_f}={t \over \tau_f}
\left(1,{ {\bf x} \over t} \right).
\label{hubumu}
\end{equation}
Note that the physical interpretation of the parameter $\tau_f$ is now slightly different from that used in the previous Section.

The calculation of the volume element of the freeze-out hypersurface shows that it is proportional to $u^\mu$, as in the model of Siemens and Rasmussen~\cite{Siemens:1978pb}, so we may write
\begin{eqnarray}
  p_\mu u^\mu		&=& \frac{1}{\tau_f} \left (\sqrt{\tau_f^2 + \rho^2}\,\, m_T \cosh(Y_s - Y) - \rho\,\, p_T \cos(\phi - \phi_p) \right)
\label{Cr-pu}
\end{eqnarray}
and
\begin{eqnarray}
  d\Sigma_\mu p^\mu 	&=&  \tau_f\, \rho\,\, p_\mu u^\mu.
\label{Cr-Sp}
\end{eqnarray}
\subsection{Resonance decays}
The two- and three-body resonance decays are implemented in exactly the same manner as in the original
version of the code~\cite{Kisiel:2005hn}. The format of the tables from SHARE \cite{Torrieri:2004zz} is presented in Appendix  \ref{AppA}.

\begin{figure}[tbh]
\begin{center}
\subfigure{\includegraphics[width=7cm]{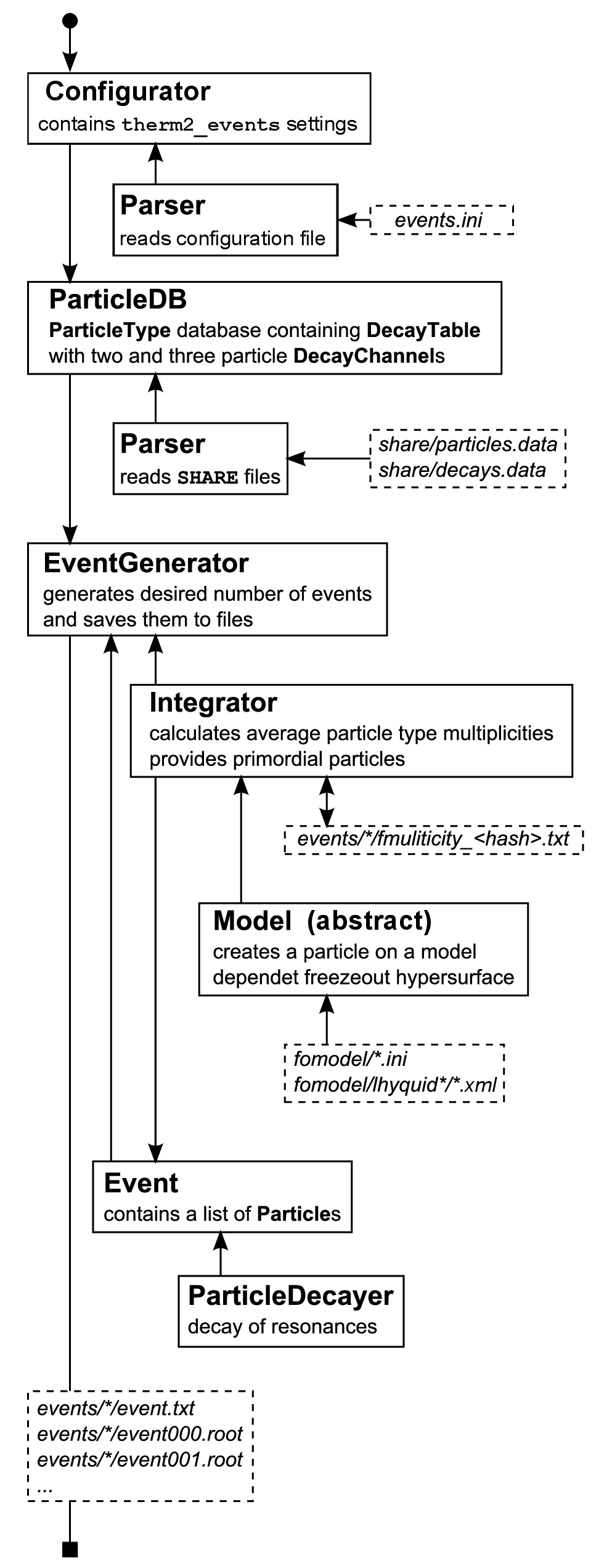}}
\end{center}
\caption{Visual concept of the {\tt therm2\_events} program. Solid line boxes represent classes or a groups of classes with their names in bold font. Dashed rectangular boxes represent input or output files. The arrows symbolize the communication between classes. 
\label{files:block_events}}
\end{figure}

\section{Structure of THERMINATOR 2}

{\tt THERMINATOR} is written in {\tt C++} and uses the CERN {\tt ROOT} environment -- {\em An object-oriented
data analysis framework}~\cite{root}.

\subsection{Modules and their functions}
\begin{figure}[tbh]
\begin{center}
\includegraphics[width=10.5cm]{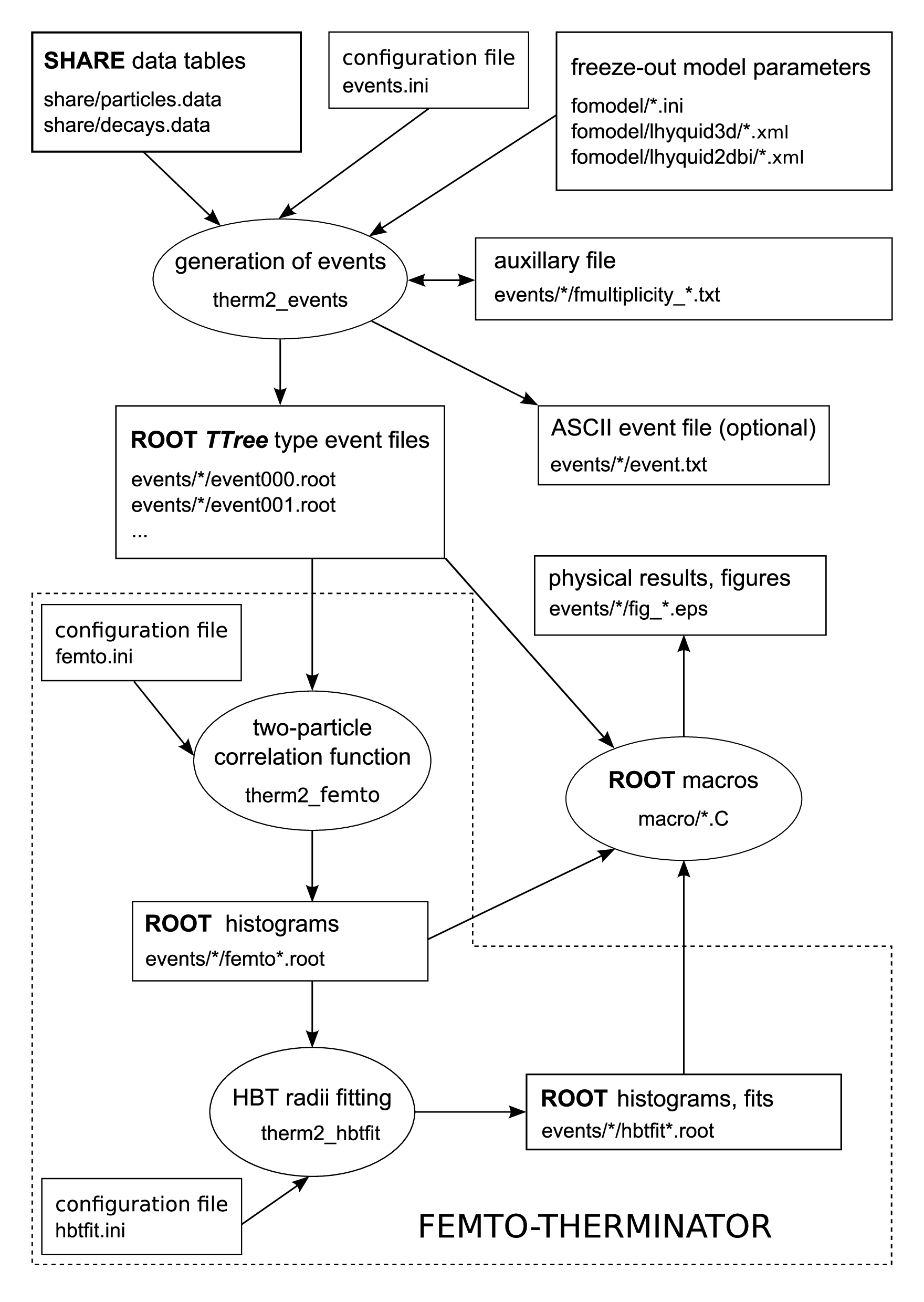}
\end{center}
\caption{Organization of input and output files. Rectangular boxes represent files, ovals correspond to codes or scripts. 
The arrows indicate reading or writing. The part enclosed in the dashed line corresponds to the {\tt FEMTO-THERMINATOR} package.
\label{files:block}}
\end{figure}

The code consists of several modules performing subsequent tasks. 
First, the information on particles and their decays is read by {\tt Parcer.cxx} and stored by {\tt ParticleDB.cxx}. Then {\tt EventGenerator.cxx} conducts the generation of events, which consists of several tasks. {\tt Integrator.cxx} computes the average multiplicity of primordial particles corresponding to a given model and parameters. Next, {\tt Event.cxx} generates the event, first producing the primordial particles with the multiplicity given by the Poisson distribution from {\tt Integrator.cxx}, and then calling {\tt ParticleDacayer.cxx} to carry out the decays of resonances. Finally, the event is written to the output file. The corresponding block structure is shown in Fig.~\ref{files:block_events}. The organization of the 
input and output files is shown in Fig.~\ref{files:block}.
\subsection{File structure}
\begin{figure}[!h]
\begin{center}
  \includegraphics[width=3.5cm]{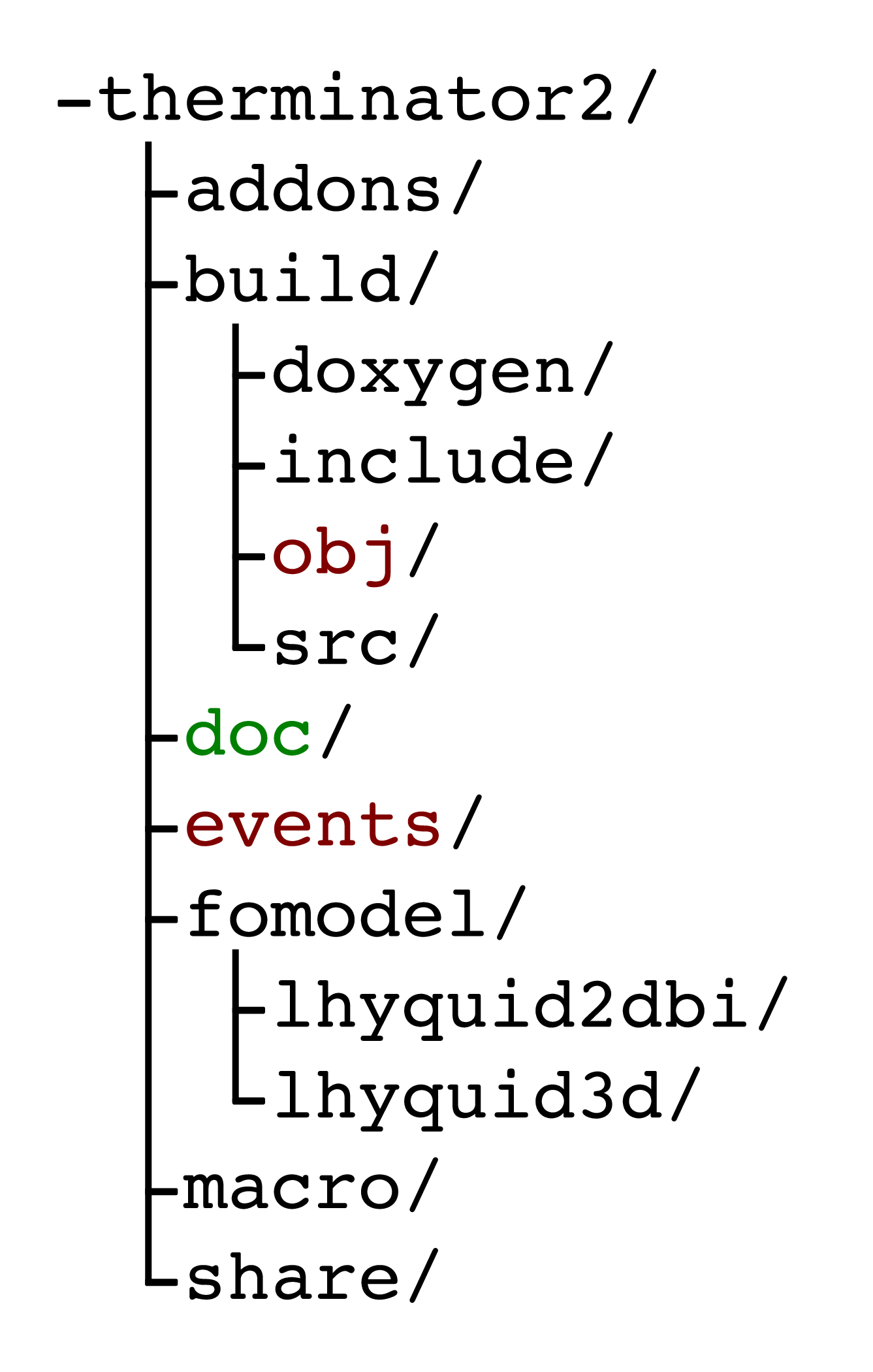}
\end{center}
\caption{Directory structure of the {\tt THERMINATOR 2} package.
\label{files:dir_struct}}
\end{figure}

After unpacking the {\tt THERMINATOR 2} package in the directory {\tt therminator2} one obtains the directory structure presented in Fig.\ref{files:dir_struct}.
The main directory of {\tt THERMINATOR 2} contains the following files:
\begin{itemize}
  \item[]{\tt Doxygen} -- Doxygen \cite{doxygenweb} configuration,
  \item[]{\tt events.ini} -- {\tt THERMINATOR 2} configuration used by {\tt therm2\_events},
  \item[]{\tt femto.ini} -- {\tt THERMINATOR 2} configuration used by {\tt therm2\_femto},
  \item[]{\tt hbtfit.ini} -- {\tt THERMINATOR 2} configuration used by {\tt therm2\_hbtfit},
  \item[]{\tt Makefile} -- make configuration,
  \item[]{\tt manual.html} -- reference manual in HTML format (created with {\tt make doc}),
                                      redirection to doc/html/index.html,
  \item[]{\tt manual.pdf} -- reference manual in PDF format (created with {\tt make doc}),
                                     symbolic link to doc/latex/refman.pdf,
  \item[]{\tt runtorque.sh} -- bash shell script for TORQUE \cite{torqueweb},
  \item[]{\tt runall.sh} -- bash shell script for the full calculation,
  \item[]{\tt runevents.sh} -- bash shell script for generating events,
  \item[]{\tt runfigure.sh} -- bash shell script for generating figures,
  \item[]{\tt runhbt.sh} -- bash shell script for the femtoscopic analysis,
  \item[]{\tt therm2\_events} -- program generating events (created with {\tt make}),
  \item[]{\tt therm2\_femto} -- program for two-particle correlations  (created with {\tt make}),
  \item[]{\tt therm2\_hbtfit} -- program for HBT analysis (created with {\tt make}).
\end{itemize}

The other directories are:
\begin{flushleft}
  \tt therminator2/build/doxygen/
\end{flushleft}
\vspace{-4mm}
-- contains Doxygen input files used to generate the reference manual,
\begin{flushleft}
\noindent
  \tt therminator2/build/include/
\end{flushleft}
\vspace{-4mm}
-- contains header files ({\tt *.h}) with the declarations of classes and structures,  as well as the general declaration file {\tt THGlobal.h},
\begin{flushleft}
  \tt therminator2/build/obj/
\end{flushleft}
\vspace{-4mm}
-- contains the compiled source files ({\tt *.o}) and the binaries {\tt therm2\_events} and {\tt therm2\_femto} (created with {\tt make}),
\begin{flushleft}
  \tt therminator2/build/src/
\end{flushleft}
\vspace{-4mm}
-- contains the source code files for the classes ({*.cxx}) and the programs {\tt therm2\_events.cxx} and {\tt therm2\_femto.cxx},
\begin{flushleft}
  \tt therminator2/doc/
\end{flushleft}
\vspace{-4mm}
-- contains the Doxygen reference manuals in the HTML and LaTeX formats (created with {\tt make doc}), 
\begin{flushleft}
  \tt therminator2/events/
\end{flushleft}
\vspace{-4mm}
-- contains the generated event files, figures, correlation function files, etc. (the contents of the {\tt events/} directory and the subdirectory structure depends on the selected model and its settings).

The default subdirectories for each model are as follows:
\begin{itemize}
  \item[]{\tt therminator2/event/krakow/} -- the Krakow Single Freeze-Out model,
  \item[]{\tt therminator2/event/blastwave/} -- the Blast-Wave model,
  \item[]{\tt therminator2/event/bwa/} -- the Blast-Wave A-class model,
  \item[]{\tt therminator2/event/lhyquid2dbi-*/}
  -- the Lhyquid 2+1D Boost-Invariant hydro,
  \item[]{\tt therminator2/event/lhyquid3d/}\footnote{{\tt THERMINATOR 2} distribution does not include hypersurfaces for this case. 
   They can be downloaded from the project homepage {\tt http://therminator2.ifj.edu.pl}.} -- the Lhyquid 3+1D hydro.
\end{itemize}

The files stored in the {\tt events/*/} subdirectory:
\begin{itemize}
  \item[]{\tt event*.root} -- event file,
  \item[]{\tt fmultiplicity\_*.txt} -- primordial particle multiplicity,
  \item[]{\tt fig\_*.eps} -- figure in EPS format,
  \item[]{\tt fig\_*.png} -- figure in PNG format,
  \item[]{\tt fig\_*.xml} -- figure data in XML format,
  \item[]{\tt fig\_*.C} -- {\tt ROOT} macro to generate a figure,
  \item[]{\tt femto*.root} -- two particle correlation function,
  \item[]{\tt hbtfit*.root} -- fit to the two particle correlation function,
  \item[]{\tt events.ini} -- event generator configuration file,
  \item[]{\tt femto.ini} -- two-particle correlation configuration file,
  \item[]{\tt hbtfit.ini} -- HBT fit configuration file,
  \item[]{\tt model\_config.ini} -- freeze-out model configuration file (the file name depends on the chosen model, e.g. {\tt lhyquid2dbi.ini}).
\end{itemize}

File naming conventions are as follows: 
The {\tt event*.root} files are always stored in a sequence beginning with {\tt event000.root}, and subsequently 
with the file counter incremented by one. The {\tt fmultiplicity\_*.txt} file has an eight-character hex-base number representing the CRC-32 identification number calculated from the model parameters, e.g.,  {\tt fmultiplicity\_10D57D490.txt} for the default case of RHIC with the centrality 20-30\%. The {\tt fig\_*.*} files have the name inherited from the corresponding ROOT macros (with the word ``figure'' 
shortened to ``fig''). The {\tt femto*.root} and {\tt fmfit*.root} filenames depend on the particle pair type, average pair momentum, 
and the inclusion of particles form resonances, e.g., {\tt femtopipi0200a.root} for the 
pion-pion pairs with average $k_T$ = 200 MeV, and with resonance decays idicated with ``a''.

\begin{flushleft}
  \tt therminator2/fomodel/
\end{flushleft}
\vspace{-4mm}
-- contains the freeze-out model configuration files ({\tt*.ini}) and freeze-out hypersurfaces data in the XML format ({*.xml}). The five freeze-out model configuration files are supplied:
\begin{itemize}
  \item[]{\tt blastwave.ini} -- for Blast-Wave model,
  \item[]{\tt bwa.ini} -- for modified Blast-Wave A-Class model,
  \item[]{\tt krakow.ini} -- for Krakow Single Freeze-out model,
  \item[]{\tt lhyquid2dbi.ini} -- for 2+1 boost-invariant hydrodynamics,
  \item[]{\tt lhyquid3d.ini} -- for 3+1 hydrodynamics.
\end{itemize}
The freeze-out hypersurface input files for RHIC at $\sqrt{s_{NN}} = 200$~GeV and various centralities have the following names:
{\scriptsize 
\begin{itemize}
  \item[]{\tt lhyquid2dbi/RHICAuAu200c0005Ti500ti025Tf145.xml} (for $c$ = 0--5 \%),
  \item[]{\tt lhyquid2dbi/RHICAuAu200c0510Ti491ti025Tf145.xml} (5--10 \%),
  \item[]{\tt lhyquid2dbi/RHICAuAu200c1020Ti476ti025Tf145.xml} (10--20 \%),
  \item[]{\tt lhyquid2dbi/RHICAuAu200c2030Ti455ti025Tf145.xml} (20--30 \%),
  \item[]{\tt lhyquid2dbi/RHICAuAu200c3040Ti429ti025Tf145.xml} (30--40 \%),
  \item[]{\tt lhyquid2dbi/RHICAuAu200c4050Ti398ti025Tf145.xml} (40--50 \%),
  \item[]{\tt lhyquid2dbi/RHICAuAu200c5060Ti354ti025Tf145.xml} (50--60 \%),
  \item[]{\tt lhyquid2dbi/RHICAuAu200c6070Ti279ti025Tf145.xml} (60--70 \%).
\end{itemize}
}
The name encodes the colliding nuclei, the collision energy (in MeV), the centrality class, the initial central 
temperature Ti (in MeV), the starting time used in hydrodynamics ti (in fm/c), and the freeze-out temperature Tf (in MeV).   
In an analogous way, the freeze-out hypersurface input files for the LHC collisions at $\sqrt{s_{NN}} = 5.5$~TeV and various centralities are:
{\scriptsize
\begin{itemize}
  \item[]{\tt lhyquid2dbi/LHCPbPb5500c0005Ti500ti100Tf145.xml} (for $c$ = 0--5 \%),
  \item[]{\tt lhyquid2dbi/LHCPbPb5500c0510Ti500ti100Tf145.xml} (5--10 \%),
  \item[]{\tt lhyquid2dbi/LHCPbPb5500c1020Ti500ti100Tf145.xml} (10--20 \%),
  \item[]{\tt lhyquid2dbi/LHCPbPb5500c2030Ti500ti100Tf145.xml} (20--30 \%),
  \item[]{\tt lhyquid2dbi/LHCPbPb5500c3040Ti500ti100Tf145.xml} (30--40 \%),
  \item[]{\tt lhyquid2dbi/LHCPbPb5500c4050Ti500ti100Tf145.xml} (40--50 \%),
  \item[]{\tt lhyquid2dbi/LHCPbPb5500c5060Ti500ti100Tf145.xml} (50--60 \%),
  \item[]{\tt lhyquid2dbi/LHCPbPb5500c6070Ti500ti100Tf145.xml} (60--70 \%).
\end{itemize}
}

\begin{flushleft}
  \tt therminator2/macro/
\end{flushleft}
\vspace{-4mm}
-- contains the ROOT macro files for generating figures in the event analysis:
\begin{itemize}
  \item[]{\tt figure\_disteta.C} --  distribution in pseudorapidity,
  \item[]{\tt figure\_distmt.C} -- distribution in transverse mass,
  \item[]{\tt figure\_distpt.C} -- distribution in transverse momentum,
  \item[]{\tt figure\_distpt\_exotic.C} --  distribution in transverse momentum of less abundant particles (e.g., those with strangeness),
  \item[]{\tt figure\_distpt\_pion.C} -- pions' anatomy in transverse momentum (contributions from individual resonance decays to pion spectra),
  \item[]{\tt figure\_distpt\_proton.C} -- protons' anatomy in transverse momentum (contributions from individual resonance decays to proton spectra),
  \item[]{\tt figure\_distrap.C} -- distribution in rapidity,
  \item[]{\tt figure\_distrap\_prim.C} -- primordial-particle distribution in rapidity,
  \item[]{\tt figure\_hsret.C} -- primordial-particle distribution in $ (\rho, \eta, \tau) $,
  \item[]{\tt figure\_hsxyt.C} -- primordial-particle distribution in $ (x, y, \tau) $,
  \item[]{\tt figure\_xpx.C} -- position-momentum correlation, $ x-p_x $,
  \item[]{\tt figure\_v2pt.C} -- differential elliptic flow at midrapidity,
  \item[]{\tt figure\_v4pt.C} -- differential $v_4$ coefficient at midrapidity,
  \item[]{\tt figure\_v2rap.C} -- integrated elliptic flow vs. rapidity.
\end{itemize}
Other ROOT macros used by the {\tt figure*.C} files are:
\begin{itemize}
  \item[]{\tt events2chain.C} -- reads event files,
  \item[]{\tt hist2xml.C} -- converts histogram data into the XML format,
  \item[]{\tt model2legend.C} -- creates figure legends displaying the model parameters.
\end{itemize}

\begin{flushleft}
  \tt therminator2/share
\end{flushleft}
\vspace{-4mm}
-- contains the SHARE \cite{Torrieri:2004zz} database files:
\begin{itemize}
  \item[]{\tt decays.data} -- particle decay channels,
  \item[]{\tt particles.data} -- particle properties.
\end{itemize}
\begin{flushleft}
  \tt therminator2/addons
\end{flushleft}
\vspace{-4mm}
-- contains an auxiliary program (estimate of the elastic rescattering):
\begin{itemize}
  \item[]{\tt coll.mk} -- makefile,
  \item[]{\tt collcount.cxx} -- program estimating elastic collisions per pion.
\end{itemize}

\begin{table}[tbh]
 \caption{The contents of the file {\tt events.ini} \label{evtab}}
{\scriptsize
\begin{verbatim}
# ----------------------|-------------------------------|
#    FreezeOutModel     |     model parameter file      |
# ----------------------|-------------------------------|
#  KrakowSFO            | ./fomodel/cracow.ini          |
#  BlastWave            | ./fomodel/blastwave.ini       |
#  BWAVT                | ./fomodel/bwa.ini             |
#  BWAVTDelay           | ./fomodel/bwa.ini             |
#  BWAVLinear           | ./fomodel/bwa.ini             |
#  BWAVLinearDelay      | ./fomodel/bwa.ini             |
#  BWAVLinearFormation  | ./fomodel/bwa.ini             |
#  Lhyquid3D            | ./fomodel/lhyquid3d.ini       |
#  Lhyquid2DBI          | ./fomodel/lhyquid2dbi.ini     |
# ----------------------|-------------------------------|

# Name of the freeze-out model
# available: see table above
# default:	Lhyquid2DBI
FreezeOutModel = Lhyquid2DBI

# Custom freeze-out model ini file [not used by default]
# default:
# FreezeOutModelINI = ./fomodel/lhyquid2dbi.ini

[Event]
# Number of events to generate 
# default:	5000
NumberOfEvents = 5000

# Event output file format
# available:	root, root&text, text
# default:	root
EventFileType = root

[Primordial]
# Distribution of primordial particles multiplicity
# available:	Poisson
# default:	Poisson
MultiplicityDistribution = Poisson

# Number of samples used in determination of primordial multiplicity and max. integrand value
# default:	5000000
IntegrateSamples = 5000000

[Random]
# Start each event with a new random seed taken from current time (1) or do a constant seed (0)
# default:	1
Randomize = 1

[Directories]
# Directory with SHARE input files
# default:	./share/
ShareDir = ./share/

# Directory with Freeze-Out Model parameter files
# default:	./fomodels/
FreezeOutDir = ./fomodel/

# Directory with ROOT macro files *.C
# default:	./macro/
MacroDir = ./macro/

# Directory to write the events
# default:	./events/
EventDir = ./events/

[Logging]
# Log file - save information on number of events runed, destination and time
# default:	therminator.log
LogFile = therminator.log
\end{verbatim}
}
\end{table}

\subsection{{\tt *.ini} files}
The main configuration file {\tt events.ini} (Table \ref{evtab}) is used by {\tt therm2\_events} program to select the model, number of generated events, type of output, file and directory names, etc. Depending on the value of {\tt FreezeOutModel}, the program reads additional information from an appropriate model settings file {\tt *.ini}. The file is located in the {\it FreezeOutDir} directory.

The model configuration file (Table \ref{tabmod}) contains the parameters of a particular model, e.g., the configuration file for the default {\bf FreezeOutModel = Lhyquid2DBI} is {\tt fomodel/lhyquid2dbi.ini}.

The variable {\tt IntegrateSamples}, used to obtain the integrand ranges and the particle average multiplicities, 
is set to a large value (5000000 by default). This leads to accurate results, but increases the running time 
of the first stage of {\tt therm2\_events}. The user may decrease this value, however, care is needed in the inspection 
of results. A very large value of {\tt NumberOfEvents}, above 20000, is needed to achieve stable HBT results. A low value (about 1000) is 
sufficient to obtain the $p_T$ spectra. Of course, the running time is proportional to {\tt NumberOfEvents}.

\begin{table}[tbh]
 \caption{Parameters of {\tt fomodel/lhyquid2dbi.ini} \label{tabmod}}
{\scriptsize
\begin{verbatim}
#########################################################
#       Lhyquid 2+1 Boost Invariant                     #
#########################################################

# Rapidity range
# default: 4.0
RapPRange = 4.0

# Spatial rapidity range
# default: 8.0
RapSRange = 8.0

# location of the hydro-code output containing the definition of the hypersurface,
# velocity profile, thermodynamic properties etc.
# default: lhyquid2dbi/RHICAuAu200c2030Ti455ti025Tf145.xml
FreezeFile = lhyquid2dbi/RHICAuAu200c0005Ti500ti025Tf145.xml

# subdirectory to store events of this model.
# If option is hashed the EventSubDir will be made form FreezeFile name (replace '/' with '-' 
# and remove ".xml")
# default: lhyquid2dbi-RHICAuAu200c2030Ti455ti025Tf145/
#EventSubDir = lhyquid2dbi-RHICAuAu200c2030Ti455ti025Tf145/
\end{verbatim}
}
\end{table}

\section{Storage of events}
\label{sect:stor}

The particles are stored in the computer memory as collections corresponding to one event. Once the whole event is generated (all primordial particles are produced and all unstable particles have been sequentially decayed), the event is written down to the output file and erased from the memory. There are two types of output: plain ASCII text and {\tt ROOT} file.

\subsection{Plain text output file}
The plain text output has the advantage of having the complete information and being easily readable by a human-being or a computer program. It is, however, not efficient in terms of disk-space usage and data-access time. The structure of the plain text file is the following:
\begin{itemize}
  \item The first two lines form the file header containing a description of the EVENT\_ENTRY, see further in text.
  \item Each event in the output file consists of an event header containing the EVENT\_ID (a unique number generated with the CRC-32 algorithm from the system time and event sequence number) and number of particles in the event.
  \item The lines that follow the event header are the particle entries. Each particle is represented by a single line.
\end{itemize}
An example of a few lines of the test output:
{\scriptsize
\begin{verbatim}
# THERMINATOR 2 text output
#<EVENT_ENTRY>eid fathereid pid fatherpid rootpid decayed mass e px py pz t x y z</EVENT_ENTRY>
#<EVENT_ID>0x65679F6</EVENT_ID>
#<NO_OF_PARTICLES>9684</NO_OF_PARTICLES>
0 -1 9001 9001 9001 0 2.350000e+00 5.423351e+00 5.105272e-01 8.400604e-01 -4.787891e+00
    3.012708e+01 3.797786e+00 1.937673e+00 -2.789553e+01
1 -1 5218 5218 5218 1 2.250000e+00 5.035699e+00 1.177745e+00 -3.007674e-01 -4.337997e+00
    2.272015e+01 2.773765e+00 -1.740429e+00 -1.985397e+01
\end{verbatim}
}
The particle event entry columns contain the following information:
{
\begin{itemize}
  \item {\bf eid}, the particle sequence number in the event,
  \item {\bf fathereid}, the sequence number of the parent, 
  \item {\bf pid}, Particle Data Group (PDG) identification number, 
  \item {\bf fatherpid}, PDG identification number of the parent, 
  \item {\bf rootpid}, PDG identification number of the original primordial particle starting the decay sequence, 
  \item the decay flag (1 if particle has decayed, 0 if not),
  \item mass and components of the four momentum: {\bf e}, {\bf px}, {\bf py}, and {\bf pz} (in GeV),
  \item space-time coordinates of the creation point: {\bf t}, {\bf x}, {\bf y} and {\bf z} (in fm/c).
\end{itemize}
}
The information on {\bf eid}, {\bf fathereid}, together with {\bf pid} and {\bf fathereid} can be used to trace back the resonance cascade from the last stable particle up to its primordial particle born on the freeze-out hypersurface. 
Particles with {\bf \hbox{fathereid = -1}} are primordial particles with {\bf pid~=~fathereid~=~rootpid}. 
\subsection{{\tt ROOT} file}
The information stored in the ROOT file is slightly different from the one stored in the text file and it is grouped in three separate {\tt TTree} type objects.

The tree called {\bf particles} consists of one branch named {\bf particle}. The branch is filled with particle entries (leaves) 
-- one leaf per particle. This tree stores information on particles from 500 events (by default). The particle leaf consist of a class called {\bf ParticleCoor} that stores the following particle properties:
{\scriptsize
\begin{center}
\begin{tabular}[!h]{cll}
  data type	& name 		& description \\
\hline
  Float\_t	& mass		& mass\\
  Float\_t	& t, x,  y,  z	& space-time coordinates (in fm/c)\\
  Float\_t	& e, px, py, pz	& energy and momentum coordinates (in GeV)\\
  Int\_t	& decayed	& decay flag\\
  Int\_t	& pid		& PDG identification number\\
  Int\_t	& fatherpid	& parent PDG number\\
  Int\_t	& rootpid	& root (primordial) particle PDG number\\
  Int\_t	& eid		& sequence number in the event\\
  Int\_t	& fathereid	& parent sequence number in the event\\
  UInt\_t	& eventid	& CRC-32 unique id of the event\\
\end{tabular}
\end{center}
}
The data types used here are the ROOT data types. 

The events tree called {\bf events} consists of one branch named {\bf event}. The branch is filled with event entries (leaves) -- 
one leaf per event, by default 500 entries per file. The event leaf consist of a structure called {\bf StructEvent} that stores the information on the number of particles in each event and the event id number:
{\scriptsize
\begin{center}
\begin{tabular}[!h]{cll}
  data type	& name 		& description \\
\hline
  UInt\_t	& eventID	& unique ID of the event generated with CRC-32 algorithm\\
  UInt\_t	& entries	& number of particle entries in this event\\
  UInt\_t	& entriesprev	& number of previous entries, 0 by default.\\
\end{tabular}
\end{center}
}
The {\bf entriesprev} variable is filled with 0 and is not used at this stage. 
It is used in the classes that read a sequence of event files for analysis.

The parameters tree called {\bf parameters} consists of eight branches that store the following information :
{\scriptsize
\begin{center}
\begin{tabular}[!h]{lcl}
branch name	& data type	& description\\
\hline
IntegrateSample	& UInt\_t	& number of Monte-Carlo integration samples\\
Randomize	& UInt\_t	& random seed switch (1-random seed, 0-constant seed)\\
TimeStamp	& Char\_t	& date and time of this run\\
ModelID		& UInt\_t	& id number of the chosen freeze-out model\\
ModelName	& Char\_t	& name of the freeze-out model\\
ModelHash	& Char\_t	& CRC-32 hash made of the freeze-out model parameters\\
ModelDescription& Char\_t	& model parameter description\\
ModelParameters	& {\it Model\_t}	& structure containing model parameters.\\
\end{tabular}
\end{center}
}

The structure {\bf Model\_t} is the default freeze-out model structure that does not contain any parameter information. Each freeze-out model has a separate structure defined and saved to the file. The detailed information can be found in the {\tt therminator2/build/include/StructModel.h} file. As an example,  the structure for the default freeze-out model is presented as follows:
{\scriptsize
\begin{center}
\begin{tabular}[!h]{cll}
data type	& name 			& description \\
\hline
Float\_t	& RapPRange		& rapidity range\\
Float\_t	& RapSRange		& spatial rapidity range\\
Float\_t	& TauI			& initial proper time\\
Float\_t	& TempF			& freeze-out temperature\\
Float\_t	& MuB			& baryochemical potential\\
Float\_t	& MuI			& isospin potential\\
Float\_t	& MuS			& strangeness potential\\
Float\_t	& MuC			& charm potential\\
Float\_t	& CollidingEnergy	& CMS beam energy\\
Float\_t	& CentralityMin		& lower limit of centrality\\
Float\_t	& CentralityMax		& higher limit of centrality\\
Float\_t	& ImpactParameter	& impact parameter\\
Float\_t	& TempI			& initial central temperature\\
Char\_t		& DeviceName[30]	& name of the device, e.g. RHIC, LHC\\
Char\_t		& CollidingSystem[30]	& colliding nuclei, e.g, Au, Pb.\\
\end{tabular}
\end{center}
}

\section{Installation and sample runs \label{sec:installation}}

\subsection{Downloading the latest version}
{\tt THERMINATOR 2} is distributed in a form of a .tar.gz archive containing the C++ sources, scripts, configuration files, and the data files. The latest package can be downloaded from the website
\begin{itemize}
  \item[] { \tt {http://therminator2.ifj.edu.pl/}} \,\,.
\end{itemize}
Prior to compiling and running {\tt THERMINATOR 2}, the user has to install the following programs:
\begin{itemize}
    \item {\tt C++} compiler,
    \item {\tt ROOT} ver. 5.26 or later.
\end{itemize}
The latest version of the {\tt ROOT} package is available at {\tt http://root.cern.ch/}.\\
The simplest way to proceed is to follow the instructions below. In this way one first creates a directory called {\tt therminator2}. Then, the package is saved and extracted,
{\small
\begin{verbatim}
mkdir therminator2
cd therminator2/
wget therminator2.ifj.edu.pl/therminator2-latest.tar.gz
tar xzf therminator2-latest.tar.gz
\end{verbatim}
}
In place of ``latest'' the actual number appears. 
The information about the version of the package is contained in the {\tt version} file.
\subsection{Compilation}
One compiles {\tt THERMINATOR 2} with the command
{\small
\begin{verbatim}
make 
\end{verbatim}
}
After a successful compilation, the binary files are present in the main directory.
\subsection{Doxygen documentation}
The source files of {\tt THERMINATOR 2} are documented with {\bf Doxygen} to generate the web-page and the PDF files with detailed information on the classes, variables, and files. The user must have the Doxygen package installed on the system (latest version can be obtained from {\tt www.doxygen.org/}). The documentation is created with the command

{\small
\begin{verbatim}
make doc
\end{verbatim}
}
In the main distribution directory, two files are created: {\tt manual.html}, which redirects the web browser to the main documentation page, and {\tt manual.pdf}, which is a symbolic link to the PDF reference manual.

\subsection{A test run}

A test run is executed with with the script

{\small
\begin{verbatim}
./runall.sh
\end{verbatim}
}

This run displays all basic functionality of the package, including the femtoscopic analysis, producing the plots with the physical results, etc.
It corresponds to Au+Au collisions at the top RHIC energy and centrality 20-30\%. 
The 2D Boost-Invariant hydrodynamics is used as the input. By default, 5000 events are used. The test run takes several hours to complete (4~h~20~min on 
a 3~GHz processor). At the end the events and figures are contained in the {\tt events/lhyquid2dbi-RHICAuAu200c2030Ti455ti025Tf145/} directory.

\subsection{Running}

Now we describe in a greater detail the elements (codes and figure macros) used in the script of the previous subsection.  

The main code of {\tt THERMINATOR 2} is {\tt therm2\_events}, which generates 
the events. It has been set by default (see the description of the {\tt *.ini} files) 
to generate 500 events for the case of the RHIC top energy and centrality 20-30\%. 
The 2D Boost-Invariant hydrodynamics is used. In order to perform the default run, issue the command
{\small
\begin{verbatim}
./therm2_events 
\end{verbatim}
}
This creates the file {\tt eventnnn.root} in the subdirectory\\ 
{\tt therminator2/events/lhyquid2dbi-RHICAuAu200c2030Ti455ti025Tf145}\\
together with a text file {\tt fmultiplicity\_10D57D490.txt} that holds the primordial particle multiplicities.
The number identifier nnn is 000 if no event files are present in the subdirectory, whereas it is one unit larger from the 
file with highest identifier if the files are already present.

The script 
{\small
\begin{verbatim}
./runevents.sh 
\end{verbatim}
}
generates as many events as requested by the {\tt NumberOfEvents} parameter in the {\tt events.ini} file. The user may add 
more events by repeating the ./runevents.sh command.

To generate via a {\tt ROOT} macro the figure showing the $p_T$ spectra,  issue the command
{\footnotesize
\begin{verbatim}
root -x './macro/figure_distpt.C("./events/
                    lhyquid2dbi-RHICAuAu200c2030Ti455ti025Tf145/",n)' 
\end{verbatim}
}
where $n$ is the number of event files used, starting from {\tt event000.root}.
The single and double quotation marks are necessary to pass the parameters to the macro.
The macro creates two files in the events subdirectory: {\tt fig\_distpt.eps} and {\tt fig\_distpt.xml}. 
The first one is the EPS graphics file, the second one contains the data from the histogram in a form of the XML file.\\

The creation of all figures from ROOT macros can be done automatically by issuing the command
{\small
\begin{verbatim}
./runfigure.sh ./events/lhyquid2dbi-RHICAuAu200c2030Ti455ti025Tf145/
\end{verbatim}
}
All executables, i.e. binaries and shell scripts, produce help information when called with parameter {\tt -h} or {\tt --help}, for instance,
{\small
\begin{verbatim}
./therm2_events -h
./runfigure --help
\end{verbatim}
}

\subsection{Auxiliary files}

After the program {\tt therm2\_events} is started, the file {\tt event\_*.tmp} is created in the main directory of the {\tt THERMINATOR 2} distribution. This file has two lines with the information about the subdirectory with the event files generated by the current run 
and the total number of event files located in that directory, including files from the previous runs. 
The name of the temporary file ends with the shell parent process ID (PPID), i.e. the PID number of the shell script that has 
started the {\tt therm2\_events} process. The auxiliary {\tt event\_*.tmp} file is used by scripts 
in order to be able to run multiple {\tt THERMINATOR 2} processes at once, e.g., on a computer cluster.

\section{Scripts to generate physical results}

We provide several {\tt ROOT} scripts that may be used to generate physical results. These scripts read the chain of {\tt event*.root} files and produce figures and {\tt ROOT} figure macros. To work properly, the scripts must read from the sequence of event files generated by {\tt THERMINATOR 2}, starting from {\tt event000.root}. This means that the user should not accidentally delete event files from the middle of a sequence.

The transverse-momentum spectra are obtained by running the {\tt figure\_sp.C} script in {\tt ROOT}. To use $n$ event files, the user executes the following command (from the {\tt therminator2} directory):
{\small
\begin{verbatim}
root -x './macro/figure_distpt.C("./events/
             lhyquid2dbi-RHICAuAu200c2030Ti455ti025Tf145/",n)' 
\end{verbatim}
}
where $n$ is the requested number of the event files to be processed.

\begin{figure}[b]
\begin{center}
  \includegraphics[width=0.85\textwidth]{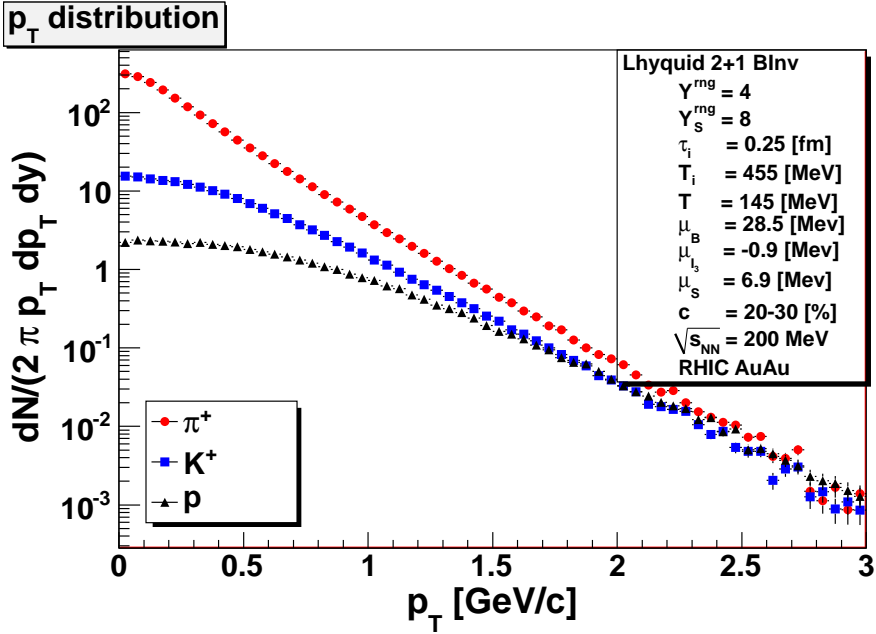}
\end{center}
\caption{\label{fig:pt} Transverse-momentum spectra of $\pi^+$, $K^+$, and protons for Au+Au collisions at $\sqrt{s_{NN}}= 200~{\rm GeV}$ and the centrality class 20-30\%, {\tt fig\_distpt.eps} (protons from $\Lambda$ decays are excluded).}
\end{figure}

For the elliptic flow coefficient, $v_2$, the user should run
{\small
\begin{verbatim}
root -x './macro/figure_v2pt.C(\"./events/
                lhyquid2dbi-RHICAuAu200c2030Ti455ti025Tf145/",n)' 
\end{verbatim}
}
This computes the elliptic flow coefficient, $v_2$, as a function of the transverse momentum, $p_T$, in the reaction plane set by the hydrodynamics.
\begin{figure}[tb]
\begin{center}
  \includegraphics[width=0.85\textwidth]{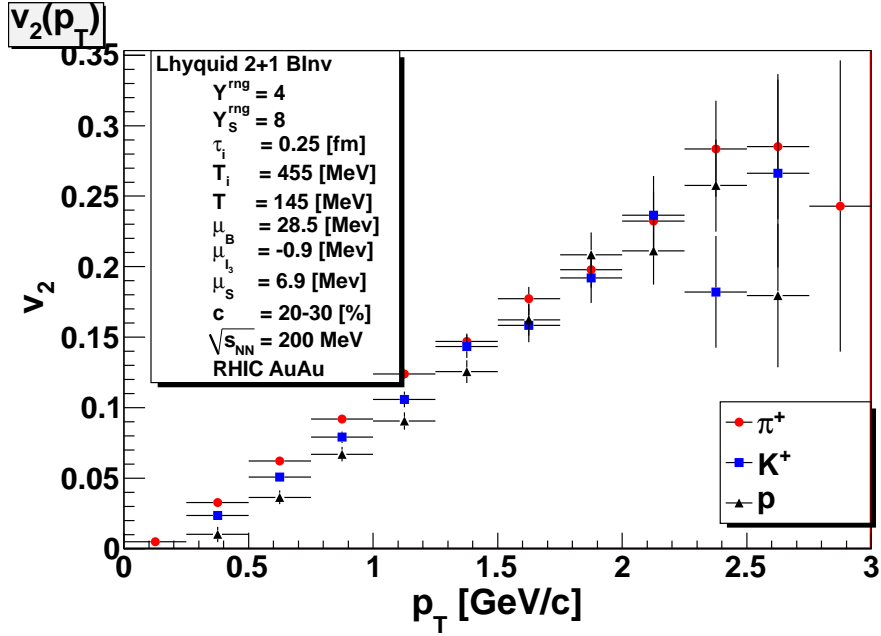} 
\end{center}
\caption{The elliptic flow coefficient, $v_2$, for $\pi^+$, $K^+$, and protons at the RHIC top energy $\sqrt{s_{NN}}= 200~{\rm GeV}$ and the centrality class 20-30\%, {\tt fig\_v2pt.eps}.}
\end{figure}
The script for the fourth-order Fourier flow coefficient, $v_4$, called {\tt figure\_v4pt.C}, is run analogously.

\section{HBT analysis \label{sec:hbt}}

This section describes the new functionality of {\tt THERMINATOR 2}, allowing to compute the 
HBT correlation radii with the two-particle method.

\subsection{Method}

The method of carrying out the femtoscopic analysis is the same as described in detail in Ref.~\cite{Kisiel:2006is}.
For integrity of this paper, we present below the main points.

Consider the two-particle distribution expressed by the two-particle emission function, 
\begin{eqnarray}
W_2({\vec p}_1, {\vec p}_2) = E_{p_1} E_{p_2} \frac {dN} {d^3 p_1 d^3 p_2}  \int S(x_1, x_2, p_1, p_2) d^4 x_1 d^4 x_2.
\label{doublespectra}
\end{eqnarray}
The correlation function is then defined as
\begin{equation}
C({\vec p}_1, {\vec p}_2) = \frac {W_2({\vec p}_1, {\vec p}_2)}
{W_1({\vec p}_1) W_1({\vec p}_2)} ,
\label{cfbyemission}
\end{equation}
where
\begin{equation}
W_1({\vec p}) = E_p {dN \over d^3p} = \int d^4x\, S(x,p) ,
\label{spectrum}
\end{equation}
with $S(x,p)$ being the source emission function.

We define the momentum difference of the particles,
\begin{equation}
q = (q_0, {\vec q}) = \left(E_{p_1} - E_{p_2}, {\vec p}_1 - {\vec p}_2 \right), 
\label{vq}
\end{equation}
the sum of their momenta, 
\begin{equation}
P = (P_0, {\vec P}) = \left(E_{p_1} + E_{p_2}, {\vec p}_1 + {\vec p}_2 \right), 
\label{vP}
\end{equation}
and the average momentum of the pair, 
\begin{equation}
k = \left( k_0, {\vec k} \right) = {{1} \over {2}} 
\left( E_{p_1} + E_{p_2}, {\vec p}_1 + {\vec p}_2 \right).
\label{vK}
\end{equation}
For the particles with equal masses we use the notation 
\begin{equation}
{\vec q}_{\rm inv} = 2 {\vec k}^*. 
\label{qinv}
\end{equation}
The space and time separations of the members of the pair are $ {\vec
  r} = {\vec r}_{1} - {\vec  r}_{2}$ and $\Delta t = t_1 - t_2$. When
calculated in the Pair Rest Frame (PRF), they are denoted as ${\vec
  r}^*$ and $\Delta t^*$. Both ${\vec k}^*$ and ${\vec r}^*$ appear as
the arguments of the wave function, since
PRF is the natural reference frame for the representation of the wave-function 
structure.   

In general, the HBT analysis may be performed in any reference
frame. One determines the correlation function as a function of the
relative momentum components in the selected frame. Then the inverse
widths of the correlation functions yields the size parameters of the
system in this frame. Here we use the standard Bertsch-Pratt
decomposition \cite{Bertsch:1988db,Pratt:1986cc} of the
mean and relative three-momenta into the three components. The {\it long}
axis coincides with the beam axis, the {\it out} axis is determined by
the direction of the average transverse momentum of the pair, denoted
later by ${\vec k}_T$, and the {\it side} direction is perpendicular
to the other two axes. Following the RHIC experiments, we choose to
perform the analysis in the longitudinal co-moving system (LCMS),
which is defined as a system where $k_{\rm long}=0$. In this work the 
notation is used in which the values in PRF are denoted by an
asterisk, while the values without asterisk are defined in LCMS. 

By definition, in a Monte-Carlo method the numerical equivalent
of the integrals  (\ref{doublespectra},\ref{spectrum}) is the summation over particles
or pairs of particles generated by the Monte-Carlo procedure. The
numerical calculation of the correlation functions is done in bins,
which may be expressed with the help of the function 
\begin{eqnarray}
\delta_{\Delta}({\vec p}) = 
\left\{
\begin{array}{cc}
1 & \hbox{if}  \,\,\,  | {p_{x}} | \leq  \frac{\Delta}{2} ,| {p_{y}} | \leq  \frac{\Delta}{2},  | {p_{z}} | \leq  \frac{\Delta}{2} \\
& \\
0 & \hbox{otherwise}.
\end{array}
\right.
\label{deltadelta}
\end{eqnarray}     
Then the correlation function may be expressed as
\begin{eqnarray}
C({\vec q}, {\vec k}) = \frac{\sum\limits_{i} \sum\limits_{j \neq i} \delta_\Delta({\vec q} 
- {\vec p}_i + {\vec p}_j ) \delta_\Delta({\vec k} - \frac{1}{2}({\vec p}_i + {\vec p}_j) )
|\Psi({\vec k}^{*}, {\vec r}^{*}) |^2} 
{\sum\limits_i \sum\limits_j \delta_\Delta({\vec  q} - {\vec p}_i + {\vec p}_j ) 
\delta_\Delta({\vec  k} - \frac{1}{2}({\vec p}_i + {\vec p}_j ))}. \nonumber \\
\label{cfbysum}
\end{eqnarray}

Various analyses of the HBT correlations use different approximations
for the full pionic wave function, $\Psi$. In the non-interacting system,
or in the interacting but non-relativistic case, the motion of the
center of mass can be separated and one deals with the relative motion
only. The simplest relative wave function ignores all dynamical
interactions and has the form 
\begin{equation}
\Psi = \frac{1} {\sqrt{2}} (e^{i {\vec k}^* {\vec r}^*} 
+ e^{-i {\vec k}^* {\vec r}^*}),
\label{psiq}
\end{equation}
where symmetrization over the two identical particles has been
performed. Therefore
$|\Psi|^2 = 1 + \cos\left(2 {\vec k}^* {\vec r}^*\right)$. 

Correlation functions calculated according to (\ref{cfbysum}) represent the ideal Bose-Einstein correlation
functions. They are also very useful in model studies, because
they can be calculated analytically for simple Gaussian emission
functions. 

In addition, in the applied method {\em mixing} of particles is introduced when evaluating the correlation functions. The purpose
of generating mixed events 
is to break all correlations other than those originating from femtoscopy, for instance the correlations 
due to resonances. The details are explained carefully in Sect.~V of \cite{Kisiel:2009eh}. 
For the case of identical charged particles this mixing technique is innocuous, as there are no resonances 
producing two identical same-charge particles. 
In addition, the mixing method increases the 
statistics, which is highly desirable in our simulations. The calculation without mixing is not implemented in the present code.

The inclusion of Coulomb effects is described in~\cite{Kisiel:2006is}.


\subsection{Calculation of correlation functions extraction of the HBT radii}
\label{sect:getrad}

The correlation functions are obtained through a
numerical implementation of Eqs. (\ref{cfbysum}) and (\ref{psiq}) (no
Coulomb effects). As described in the previous sections, particles generated by {\tt THERMINATOR 2} are grouped
into events, as in experiment. In each event every charged pion is
combined with every other pion of the same charge. For each pion pair,
$|\Psi|^2$ is calculated and added to the numerator of
Eq. (\ref{cfbysum}) in a bin corresponding to the pair's $q_{\rm out}$,
$q_{\rm side}$ and $q_{\rm long}$. At the same time, 1 is added to the
denominator of Eq. (\ref{cfbysum}) in the corresponding bin. The
resulting ratio yields the correlation function.   

By making a proper selection of single pions and pairs of pions one
may study the correlation functions as functions of various
variables. For instance, taking into account the pairs of particles
within a certain total momentum range only, one obtains
femtoscopic information for a given $k_T$. 

Usually, the single-particle emission function is postulated to
be a static 3-dimensional ellipsoid with a Gaussian density profile 
\begin{eqnarray}
S(\vec x, \vec p) = N \exp\left(-{x_{\rm out} ^2 \over {2 R_{\rm out}^2}} -
{x_{\rm side}^2 \over {2 R_{\rm side}^2}} 
- {x_{\rm long}^2 \over {2 R_{\rm long}^2}} \right).
\label{staticsource}
\end{eqnarray}
This source function is static, i.e., it does not depend on particle
momentum.  In this case the integral leads to the well known formula: 
\begin{eqnarray}
& & C\left(k_\perp,q_{\rm out},q_{\rm side},q_{\rm long} \right) = 1 + 
\lambda \exp\left[
-R^2_{\rm out}(k_\perp) q^2_{\rm out} \right.
\nonumber \\
& & \hspace{4cm} \left.
-R^2_{\rm side}(k_\perp) q^2_{\rm side}
-R^2_{\rm long}(k_\perp) q^2_{\rm long}
\right].
\label{cfgaus}
\end{eqnarray}
The quantities $R_{\rm out}$, $R_{\rm side}$ and $R_{\rm long}$, known
as the ``HBT radii'', are the widths of the Gaussian approximation. 
It is important to emphasize that formula (\ref{cfgaus}) is commonly used to fit the
experimental data and to represent the results of the model
calculations, although the experimental or model emission functions are
frequently far from gaussians. In the simple fitting examples given in
this paper we do not address this problem. More detailed discussion of
such effects can be found in~\cite{Kisiel:2006is}.

\subsection{therm2\_femto and therm2\_fithbt programs}

The program that implements the two-particle femtoscopic formalism described
above, see Eq.~(\ref{cfbysum}), is called {\tt therm2\_femto}.
The syntax of the command is:
{\footnotesize
\begin{verbatim}
./therm2_hbt <KTBIN> <EVENT_DIR> <EVENT_FILES> [FEMTO_INI] [PPID]
\end{verbatim}
}
where the parameter {\tt KTBIN=0,1,2,3} selects the transverse-momentum bin of the pair, 
{\tt EVENT\_DIR} is the directory where the {\tt event*.root} files are stored, 
and {\tt EVENT\_FILES} is the number of the files to be taken. Parameter {\tt FEMTO\_INI} (optional) is the name configuration file. This 
parameter is by default is set to {\tt femto.ini}. The {\tt PPID} parameter (optional) is the system's process ID number
used by the shell scripts. By default it is equal to 0.

The $k_T$ bins correspond to the following ranges of the pair momentum: 
{\scriptsize
\begin{verbatim}
0 - (0.15, 0.25) GeV/c, 
1 - (0.25, 0.35) GeV/c, 
2 - (0.35, 0.45) GeV/c, 
3 - (0.45, 0.6)  GeV/c.
\end{verbatim}
}

\begin{table}[b]
\caption{Contents of the {\tt femto.ini} configuration file.}
{\scriptsize
\begin{verbatim}

[Pair]
# Type of particle pairs to correlate
# default: pion-pion, kaon-kaon
PairType = pion-pion

[Cuts]
# Particle time cut [fm]
# default:      500.0
TimeCut = 500.0

[Event]
# Number of events to mix
# default:      20
EventsToMix = 20

[Switches]
# Use only primordial particles (yes) or all particles from resonance decays (no)
# default:      no
EnableOnlyPrimordial = no

# Enable source histograms
# default:      no
EnableSourceHistograms = no

[Logging]
# Log file - save information on number of events runed, destination and time
# default:      therminator.log
LogFile = therminator.log
\end{verbatim}
}
\end{table}

The output of the {\tt therm2\_femto} code consists of the {\tt ROOT} files
{\tt femto*.root}, with histograms of the numerator and
denominator, 

which can be used to calculate the correlation
function. In addition we provide the projections of the pair emission
function $S$ on the three components of the pair relative separation
$r$ calculated in LCMS. They are saved as 2D histograms, with the
magnitude of relative momentum $q_{inv}$ at the other axis.

The {\tt therm2\_hbtfit} program is an implementation of the fitting procedure, which
uses the simplest fitting functional form (\ref{cfgaus}). 
The syntax of the commands is:
\begin{verbatim}
therm2_hbtfit <CORR_FILE> [HBTFIT_INI] [PPID]
\end{verbatim}

The first parameter {\tt CORR\_FILE} is the file containing the numerator and denominator histograms. 
The second parameter (optional) is set by default to {\tt ./fithbt.ini}. It is the name of the configuration file, 
containing the fitting parameters and switches. The third parameter (optional) is the process ID number, used by shell scripts.

\begin{table}[b]
\caption{Contents of the {\tt hbtfit.ini} file.}
{\scriptsize
\begin{verbatim}

[Norm_Fit]
# Normalization initial value
# default:      1.0
Norm = 1.0
# Fit type
# default:      fixed
NormFitType = fixed
[Lambda_Fit]
# Lambda initial value
# default:      1.0
Lambda = 1.0
# Fit type
# default:      free
LambdaFitType = free
[Rout_Fit]
# R_out initial value
# default:
Rout = 5.0
# Fit type
# default:      limit
RoutFitType = limit
# Rout limits
# default:
RoutMin = 2.0
RoutMax = 7.0
[Rside_Fit]
# Rside initial value
# default:
Rside = 5.0
# Fit type
# default:      limit
RsideFitType = limit
# Rout limits
# default:
RsideMin = 2.0
RsideMax = 7.0
[Rlong_Fit]
# Rlong initial value
# default:
Rlong = 7.0
# Fit type
# default:      limit
RlongFitType = limit
# Rout limits
# default:	
RlongMin = 4.0
RlongMax = 14.0
[QRange]
# Maximum range of q_out, q_side, q_long
#default:      0.15
MaxFitRange = 0.15
[Histograms]
# Numerator histogram name
# default:      cnuma
Numerator = cnuma
# Denominator histogram name
# default:      cdena
Denominator = cdena
[Logging]
# Log file
# default:      therminator.log
LogFile = therminator.log
\end{verbatim}
}
\end{table}

The naming convention for the parameters is as follows: The label {\tt [*\_fit]} indicates a parameter to be fitted.
The name of the parameter, e.g., {\tt Rout}, is assigned its initial value from which the fit starts. 
The {\tt *FitType} flag defines the type of fitting. If it is set to {\tt fixed}, then this value is fixed to the starting value 
in the fitting procedure. If {\tt limit} is given, the value can vary in the range given by the {\tt *Min} and {\tt *Max} parameters. 
Finally, if it is equal to {\tt free} than the value can be fitted without constraints. 

The five parameters, {\tt Norm} (set to 1), {\tt Lambda}, {\tt Rout}, {\tt Rside}, and  {\tt Rlong}, correspond to the parameters of Eq.~(\ref{cfgaus}): 
the overall normalization, the $\lambda$ parameter, and the three HBT radii ($R_{out}$, $R_{side}$, and $R_{long}$). 
The parameter  {\tt MaxFitRange} contains the range of $q$ values in which the fitting is
performed. 

The output of the {\tt therm2\_hbtfit} program is a text file {\tt hbtradii.txt}.  It contains
the values of the fitted parameters and statistical errors. Subsequent runs of the program append information to this file. 
The macro {\tt macro/figure\_hbtradii.C} reads that information and creates a plot of the HBT radii.
The {\tt therm2\_hbtfit} program also produces the output {\tt ROOT} file {\tt hbtfit*.root} which should be 
inspected in order to judge the quality of the fit. It contains the one-dimensional slices of the calculated correlation function, 
as well as the corresponding slices of the fitted function. 

The histograms are named in the following way:
\begin{verbatim}
<Type>_<Direction>_<Width>
\end{verbatim}
where {\tt Type} can be {\tt CFproj} for the correlation function slices or {\tt FITptoj} for the fitted function slices. {\tt Direction} is one of {\tt out}, {\tt side} or {\tt long}, while {\tt Width} can have a value of 1 - corresponding to a slice which has a width of 1 bin in the other two directions, 2 - a slice with the width of 4 bins and 3 - a slice with the width of 10 bins. The slices of the correlation function and the fitted function for the same direction and the same width should be plotted on top of each other in order to confirm that the fitting procedure correctly captured the width of the correlation. This is done by the macro {\tt macro/figure\_fitfemto.C}.

We warn the user that obtaining statistically stable results 
for the HBT radii requires sometimes a very large number of events, such as 20000 or even 50000. This 
requires an appropriate change of the parameter {\tt NumberOfEvents} in the {\tt events.ini} file, leading 
to long execution times. 

\begin{figure}[tb]
\begin{center}
  \includegraphics[width=\textwidth]{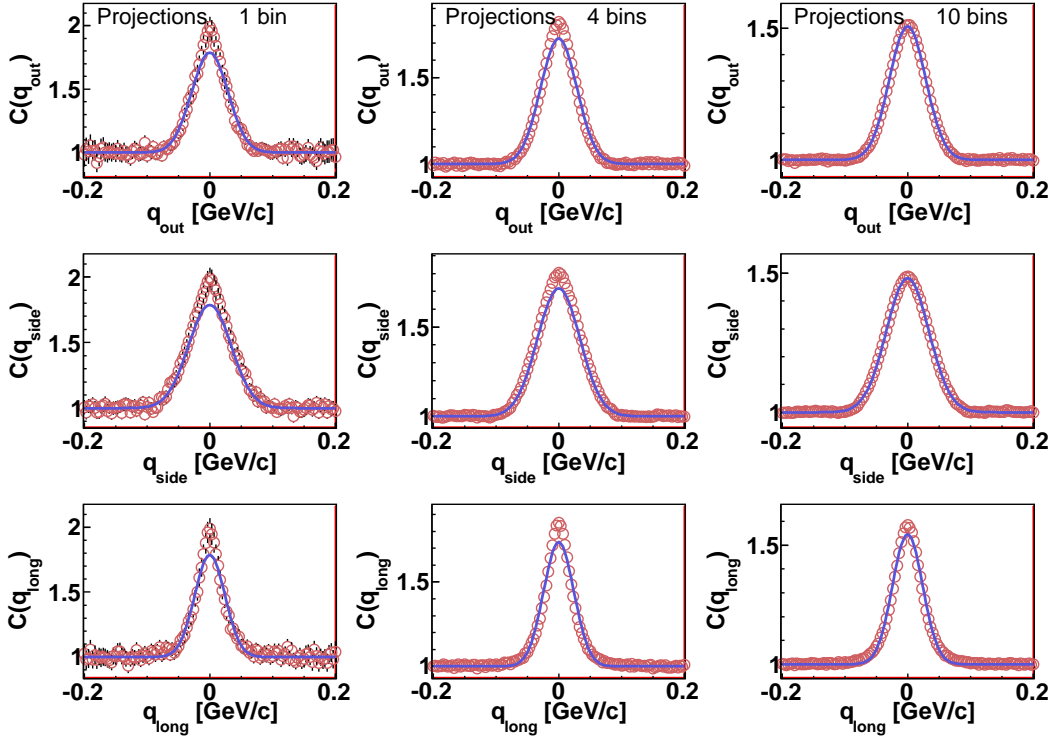} 
\end{center}
\caption{\label{fig:femtoproj}{Projections of the correlation function (symbols) calculated with the default settings for pions with $k_T$ in the (0.15,~0.25~GeV) bin. Au+Au, $\sqrt{s_{NN}}= 200~{\rm GeV}$,
centrality class 20-30\%, 50000 events (\tt fig\_cfprojpipi0a.eps}). The lines show the fit projected the same way.}
\end{figure}

\begin{figure}[tb]
\begin{center}
  \includegraphics[width=\textwidth]{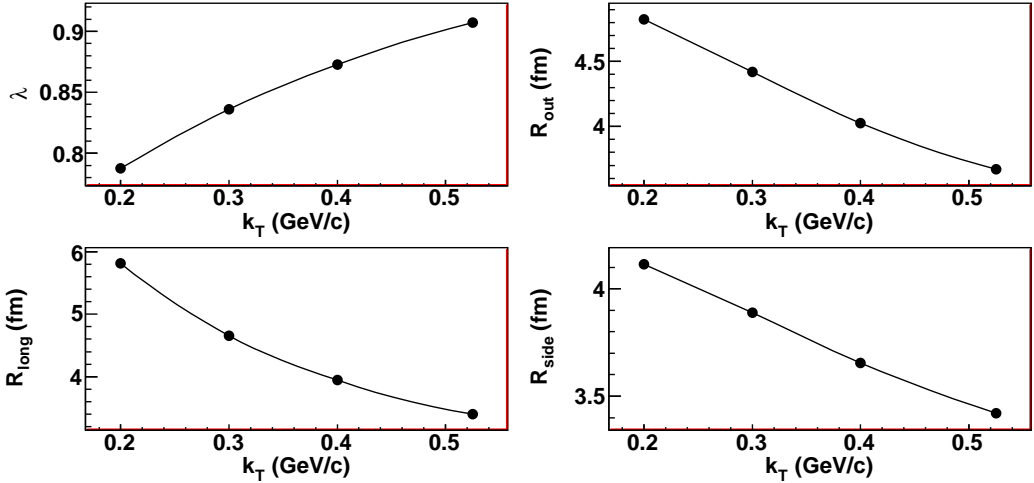} 
\end{center}
\caption{\label{fig:hbtfit} The HBT radii and $\lambda$ as a function of the average pair $k_T$ calculated with the default settings, Au+Au, $\sqrt{s_{NN}}= 200~{\rm GeV}$, centrality class 20-30\%, 50000 mixed events ({\tt fig\_hbtradii.eps}).}
\end{figure}

The running sequence of the femtoscopic analysis has been automatized in the shell script {\tt ./runhbt.sh}. The user should simply execute 
\begin{verbatim}
./runhbt.sh
\end{verbatim}
which in addition generates the physical plots. An example is presented in Fig. \ref{fig:femtoproj} and Fig. \ref{fig:hbtfit}.

\section{Yielding auxiliary results\label{sec:new}}

In this Section we present new analyses, possible with 
{\tt THERMINATOR 2}. In fact, they relate to fundamental physical
aspects of the freeze-out model assumed in our approach.

\subsection{Backflow}

The first effect concerns the so-called back-flow problem \cite{Bugaev:1996zq,Bugaev:1999wz,Bugaev:2002ch}. 
It originates from the fact that for typical freeze-out hypersurfaces generated with 
hydrodynamics, particles generated with the Cooper-Frye formula may flow back into the 
hydrodynamics domain, thus violating causality. A typical remedy is the removal 
(used also in {\tt THERMINATOR 2}) of the back-flowing particles via the constraint $p^\mu d \Sigma_\mu > 0$, where   
$d \Sigma_\mu$ is normal to the hypersurface. To see how big is the back-flow effect, we 
compute the back-flowing particles requesting that $p^\mu d \Sigma_\mu < 0$, and compare 
the result to the case without this constraint, i.e. including all particles.

To prepare the binary file of {\tt therm2\_events} generating the back-flowing
particles only, the user should run  (this feature is implemented for the {\tt Lhyquid2dbi} 
and {\tt Lhyquid3d} models only) 
{\small
\begin{verbatim}
make all BACK_FLOW=1
\end{verbatim}
}
and then execute
{\small
\begin{verbatim}
./therm2_events
\end{verbatim}
}
The user should take care of the fact that the directory specified with  {\tt EventSubDir} does not include the 
{\tt fmultiplicity\_*.txt} file, such that the multiplicities are recomputed for the present case of the back-flowing
particles.

\begin{figure}[tb]
\begin{center}
\includegraphics[width=0.85\textwidth]{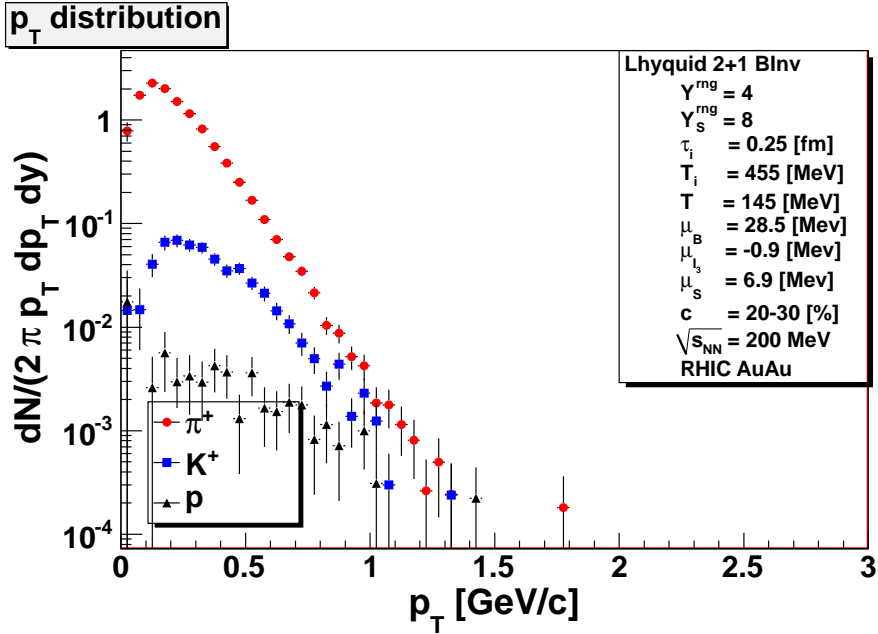} 
\end{center}
\caption{\label{fig:backflow} Same as Fig.~\ref{fig:pt} for the back-flowing particles, 5000 events.}
\end{figure}

The result is shown in Fig.~\ref{fig:backflow}. Comparing with 
Fig.~\ref{fig:pt}, we note that the back-flow effect is very small (less than 0.5\% for the pions and even less
for more massive particles) in our realistic 
calculation. This smallness is caused by large transverse velocity of the fluid, which inhibits 
the thermal motion of particles in the opposite direction \cite{Broniowski:2008vp}.

\subsection{Elastic rescattering}

The other considered effect concerns the elastic rescattering after freeze-out. In our single-freeze-out
approximation it is assumed that the chemical and thermal freeze-outs occur simultaneously. As a matter 
of fact, this is not the case, as the inelastic processes, responsible for the chemical equilibrium, have 
generally a smaller cross section than the elastic processes. In order 
to estimate how good this approximation is, one may consider a simple estimate of elastic 
rescattering based on trajectory crossings.  

Consider two particles, born at the freeze-out hypersurface at coordinates $(t_1, \vec{x}^0_1)$ and
$(t_2, \vec{x}^0_2)$. Their trajectories are 
\begin{eqnarray}
\vec{x}_1(t)= \vec{x}^0_1+(t-t_1) \vec{v}_1, \;\; \vec{x}_2(t)= \vec{x}^0_2+(t-t_2) \vec{v}_2,
\end{eqnarray}
and the closest approach occurs at 
\begin{eqnarray}
t_m= -\frac{(\vec{v}_1-\vec{v}_2)\cdot (\vec{x}^0_1-\vec{x}^0_2)}{(\vec{v}_1-\vec{v}_2)\cdot (\vec{v}_1-\vec{v}_2)}.
\end{eqnarray}
Since this time cannot be earlier than $t_1$ or $t_2$, we define $t_0={\rm max}(t_m, t_1, t_2)$. The distance squared at the closest approach 
is $d^2 = (\vec{x}_1(t_0)-\vec{x}_2(t_0))^2$. The collision occurs if, geometrically, $\pi d^2 < \sigma_{\rm el}(s)$, where 
$\sigma_{\rm el}(s)$ is the elastic cross section for the Mandelstam variable $s$. 

We have analyzed in detail the case of a charged pion scattering elastically off the other pions. The elastic 
cross section is computed with the help of the realistic parametrization of the pion-pion phase shifts from 
\cite{Colangelo:2001df}. The result is that for most-central collisions the pion scatters elastically off the other pion 
on the average 1.8 times, while for the mid-central and peripheral collisions this value drops below 1. This shows that for the 
pions the effects of elastic rescattering after freeze-out are small and our approximation is appropriate.
 
The user may carry out this analysis by running
{\small
\begin{verbatim}
cd addons
cp <dest_dir>/event000.root .
make -f coll.mk
./collcount 
\end{verbatim}
}
where {\tt <destdir>} denotes the directory with the previously generated event files for the model and centrality of interest. 

The code {\tt collcount.cxx} may also serve as a simple example of an ``off-line'' analysis with the {\tt THERMINATOR 2} events. Its 
structures (e.g., concerning the reading of events) may be directly imitated in similar applications.

\section{Conclusion}

We hope that {\tt THERMINATOR 2} will continue to be a useful and versatile tool for the heavy-ion community.
In particular, the possibility of implementing the general freeze-out conditions (read from external files)
allows the user to investigate in detail the predictions 
of the hydrodynamic approach followed with statistical hadronization (implemented in the single-freeze-out scenario). 
This approach was very successful in reproducing, in a uniform way, the RHIC data \cite{Broniowski:2008vp} and 
the LHC data at $\sqrt{s_{NN}}=2.76$~TeV~\cite{Aamodt:2010pa,Bozek:2010er,Bozek:2011wa}. It was also 
very helpful in a variety of analyses 
\cite{Broniowski:2005ae,Pratt:2005bt,Andronic:2005yp,Kisiel:2006is,Chojnacki:2006tv,Florkowski:2006mb,Biedron:2006vf,
Utyuzh:2006nw,Begun:2006uu,Utyuzh:2007ct,Brown:2007raa,Baeuchle:2007et,Vertesi:2007ki,Danielewicz:2007jn,%
Begun:2007du,Tomasik:2007gs,Chojnacki:2007rq,Bozek:2007qt,Afanasiev:2007kk,Akiba:2008zz,Broniowski:2008vp,%
Florkowski:2008cs,Lacey:2008kd,Chung:2008fu,Kisiel:2008ws,:2008fq,Tomasik:2008fq,Bozek:2008zw,:2008gk,Broniowski:2008qk,%
Broniowski:2008ee,Florkowski:2009vr,Chajecki:2009zg,Bozek:2009ty,Luo:2009sx,Bozek:2009zz,NoronhaHostler:2009tz,%
Broniowski:2009fm,Florkowski:2009rt,Bozek:2009mz,Hauer:2009nu,Ryblewski:2009hm,Broniowski:2009te,%
Videbaek:2009zy,Bozek:2009dt,Kisiel:2009iw,Kisiel:2009eh,Kisiel:2009kn,Florkowski:2009wb,Heinz:2009xj,Bozek:2009pu,Lokhtin:2009hs,NoronhaHostler:2010yc,Luo:2010by,Bozek:2010bi,Aggarwal:2010wy,Hauer:2010sv,NoronhaHostler:2010hy,Bozek:2010aj,Bilandzic:2010jr,Lizhu:2010zh,Bozek:2010vz}. 
Therefore, it has proved to be realistic and may serve as a practical tool in other physical studies or in detector modeling. 

A new part of the package, {\tt FEMTO-THERMINATOR}, carries out the femtoscopic analysis  
using the two-particle method. This is of practical importance, as 
complete analyses of the heavy-ion data typically consist of the studies of spectra, flow, and femtoscopy -- all 
are now implemented in {\tt THERMINATOR 2}.

We note that a small modification of the approach presented in this paper, accounting for viscosity effects which draw the distribution functions away from the thermal equilibrium, may be included in the straightforward way. That way the code can be used as an afterburner in the increasingly popular hydrodynamic studies with viscosity effects \cite{Muronga:2001zk,Teaney:2003kp,Baier:2006um,Baier:2006gy,Romatschke:2007mq,Chaudhuri:2006jd,Heinz:2005bw,Bozek:2007di,Bozek:2007qt,Song:2007ux,Bozek:2008bh,Song:2008hj,Luzum:2008cw,Bozek:2009ut,Bozek:2009dw,Romatschke:2009im}. Fugacities, reflecting incomplete chemical equilibrium \cite{Letessier:1998sz} of the system, can now be also incorporated. 
This allows for model studies of this phenomenon.

We have made a considerable effort to make the structure of the code clear and straightforward 
to adapt for other applications. The program is documented in detail, also via the Doxygen utility.
Numerous {\tt ROOT} macros, allowing to carry out various tasks such as the calculation and presentation of physical results,
are supplied with the package.

We foresee the following further extensions and applications of the program:

\begin{enumerate}

\item Statistical hadronization for {\em viscous} hydrodynamics -- in this case the Cooper-Frye 
formula requires a straightforward modification of the form of the distribution function $f$. 
 
\item Implementation of more advanced fitting procedures of the femtoscopic 
correlation functions, for instance imaging \cite{Brown:1997ku}. We note that even at present the 
output from {\tt FEMTO-THERMINATOR} contains all necessary information for 
such analyses (the 3-dimensional correlation functions).

\item Analysis of the triangular and multipole flow, analysis of the two-particle 
correlations other than those due to the Bose-Einstein statistics.

\item Implementation of a more general form ({\em e.g.}, negative binomial) of the 
multiplicity distributions. This will allow for a realistic modeling of event-by-event 
multiplicity fluctuations.
 
\item Conservation laws. A modification of the particle-generation algorithm might include 
(approximate) conservation of the global charges and the transverse momentum. 
 
\item Interface to other formats for the files parameterizing hypersurfaces will be developed 
according to demand of potentially  interested users.

\end{enumerate}

\section*{Acknowledgments}
We thank Piotr Bo\.zek for providing the files with hypersurfaces from the 3+1-dimensional perfect hydrodynamics. 

\newpage
\appendix

\section{Particle data files ({\tt particles.data}  and  {\tt decays.data})}
\label{AppA}

The structure of input files containing the information on the particle properties and their decays is exactly the same as in the original code~\cite{Kisiel:2005hn} --- the form inherited from {\tt SHARE} \cite{Torrieri:2004zz}. To maintain integrity,  we provide here the most basic information. The input file {\tt particles.data} contains the properties of particles such as mass, width, spin, isospin, the quark contents, and the Monte Carlo identification number.  The file has the format:

\vspace{0.1cm}

\noindent {\bf name\quad mass\quad  width\quad  spin\quad  I\quad
I3\quad  q\quad  s\quad  aq\quad  as\quad  c\quad   ac\quad  MC}

\vspace{0.1cm}

\noindent where:

\begin{description}
\setlength{\itemsep}{0.1cm}
\setlength{\labelwidth}{1.5cm}
\setlength{\itemindent}{1.1cm}
\item[name] the particle label used in the program,
\item[mass] mass in GeV,
\item[width] width in GeV,
\item[spin] spin,
\item[I] isospin,
\item[I3] 3rd component of isospin,
\item[q,\ s\hfill] number of light/strange quarks,
\item[aq,\ as\hfill] number of light/strange antiquarks,
\item[c,\ ac\hfill] number of charmed/anticharmed quarks,
\item[MC] particle's identification number, where
  available corresponding to the standard Monte Carlo particle
  identification convention \cite{Hagiwara:2002fs}.
\end{description}

\noindent The file {\tt decays.data} contains  the information on particle decay channels in the format:

\vspace{0.1cm}

\noindent {\bf Name$_{\rm parent}$  Name$_{\rm daughter1}$  Name$_{\rm
    daughter2}$  Name$_{\rm daughter3}$\footnote{
Appears for the three-body decays only.}  BR C--G?(0/1)}

\vspace{0.1cm}

\noindent where {\bf BR} denotes the branching ratio of the decay
and {\bf C--G} refers to whether the branching ratio should be multiplied by a
Clebsch--Gordan coefficient ({\bf 0}:~no, {\bf 1}: yes). Normally, we use {\bf 1} for the two-body
decays and {\bf 0} for the three-body decays, where the full (isospin-dependent)
branching ratio BR is provided. The entry BR may be used to control the feed-down from
the weak decays, in particular setting it to 0 would switch off the particular decay channel.

\section{General definitions}

Variables used to describe the properties of a particle:
{\scriptsize
\begin{center}
\begin{tabular}{l|c|l}
{\bf ParticleType.h}	& {\bf symbol} & {\bf definition} \\
\hline
mNumber			&			& particle type number\\
mName			&			& particle name\\
mMass			& $m$			& mass\\
mGamma			& $\Gamma$		& width\\
mSpin			& $s$			& spin\\
mBarionN		& $B$			& baryon number\\
mI			& $I$			& isospin\\
mI3			& $I_3$			& 3rd component of isospin\\
mStrangN		& $S$			& strangness\\
mCharmN			& $C$			& charm\\
mNq			& $n_q$			& number of $u$ and $d$ quarks\\
mNaq			& $n_{\overline{q}}$	& number of $\bar u$ and $\bar d$ quarks\\
mNs			& $n_s$			& number of $s$ quarks\\
mNas			& $n_{\overline{s}}$	& number of $\bar s$ quarks\\
mNc			& $n_c$			& number of $c$ quarks\\
mNac   			& $n_{\overline{c}}$	& number of $\bar c$ quarks\\
mPDGCode		&			& particle MC number according to PDG \cite{Hagiwara:2002fs}\\
mMaxIntegrand		&			& maximum value of the Cooper-Frye integrand\\
mMultiplicity		&			& average multiplity\\
mDecayChannelCount2	&			& number of two-body decay channels\\
mDecayChannelCount3	&			& number of three-body decay channels.
\end{tabular}
\end{center}
}

\newpage

Variables used to calculate the Cooper-Frye integrand in the {\tt Model\_*.cxx} files. See function {\tt GetIntegrand(ParticleType* aPartType)}.
{\scriptsize
\begin{center}
\begin{tabular}{l|c|l}
{\bf in program}	& {\bf symbol} & {\bf definition} \\
\hline
\hline
dSigmaP		& $p \cdot d\Sigma$	& particle's four momentum projected \\ 
            &                       & on the hypersurface element\\
PdotU		& $p \cdot u$		& particle's energy in the fluid local rest frame \\
Statistics	& $\pm1$		& $+1$ (for fermions), $-1$ (for bosons)\\
\hline
\multicolumn{3}{c}{\it space-time}\\
\hline
(Xt,\,Xx,\,Xy,\,Xz)	& $(t,x,y,z)$	& Cartesian coordinates\\
Tau		& $\tau$		& proper time\\
Rho		& $\rho$		& transverse size\\
PhiS		& $\phi$		& azimuthal angle\\
RapS		& $Y_s$			& space-time rapidity\\
\hline
\multicolumn{3}{c}{\it momentum}\\
\hline
(Pe,\,Px,\,Py,\,Pz)	& $(E,p_x,p_y,p_z)$	& Cartesian coordinates\\
Mt		& $m_T$			& transverse mass\\
Pt		& $p_T$			& transverse momentum\\
PhiP		& $\phi_p$		& transverse-momentum angle\\
RapP		& $Y$			& rapidity\\
\hline
\multicolumn{3}{c}{\it fluid element}\\
\hline
Ux, Uy		& $u_x$, $u_y$		& transverse four-velocity\\
Vt		& $v_T$			& transverse velocity\\
PhiF		& $\phi_f$		& transverse-velocity angle\\
RapF		& $Y_f$			& fluid rapidity.\\
\end{tabular}
\end{center}
}

\section{Hydro-inspired models}
Common parameters for hydro-inspired models:
{\scriptsize
\begin{center}
\begin{tabular}{l|l|c|l}
\bf ini file	& \bf in program & \bf symbol	& \bf description\\
\hline
RapPRange		& mRapPRange	& $Y^{\rm rng}$		& rapidity range\\
RapSRange		& mRapSRange	& $Y_s^{\rm rng}$	& spatial-rapidity range\\
RhoMax			& mRhoMax	& $\rho^{\rm max}$	& firecylinder transverse size\\
Temperature		& mTemp		& $T$			& freeze-out temperature\\
MuB			& mMuB		& $\mu_B$		& baryon chemical potential\\
MuI			& mMuI		& $\mu_{I_3}$		& isospin chemical potential\\
MuS			& mMuS		& $\mu_S$		& strange chemical potential\\
MuC			& mMuC		& $\mu_C$		& charm chemical potential. \\
\end{tabular}
\end{center}
}
Integration ranges
{\scriptsize
\begin{center}
\begin{tabular}{rcl|rcl}
  $Y_s$		& $\in$ & $\left[ - \frac{Y_s^{\rm rng}}{2}, \frac{Y_s^{\rm rng}}{2} \right]$ &
  $Y$		& $\in$ & $\left[ - \frac{ Y^{\rm rng}}{2}, \frac{ Y^{\rm rng}}{2} \right]$ \\
  $\rho$	& $\in$ & $\left[0 , \rho_{\rm max} \right]$ &
  $p_T$		& $\in$ & $\left[0 , \infty \right)$\\
  $\phi$	& $\in$ & $\left[0, 2 \pi \right]$ &
  $\phi_p$	& $\in$ & $\left[0, 2 \pi \right]$
\end{tabular}
\end{center}
}
\subsection{Cracow Single Freeze-out Model}
Model specific parameters:
{\scriptsize
\begin{center}
\begin{tabular}{l|l|c|l}
\bf fomodel/krakow.ini	& \bf in program & \bf symbol	& \bf description\\
\hline
TauC			& mTauC		& $\tau_f$		& Cracow proper time. \\
\end{tabular}
\end{center}
}
Definition of the proper time used in the Cracow model
\begin{equation}
  \tau_f = \sqrt{t^2-x^2-y^2-z^2} = \sqrt{\tau^2+\rho^2}.
\end{equation}
\subsection{Blast-Wave}
Model specific parameters:
{\scriptsize
\begin{center}
\begin{tabular}{l|l|c|l}
\bf fomodel/blastwave.ini	& \bf in program & \bf symbol	& \bf description\\
\hline
Tau			& mTau		& $\tau$		& proper time\\
VelT			& mVt		& $v_T$			& transverse velocity of the fluid element.\\
\end{tabular}
\end{center}
}
\subsection{Blast-Wave A-Class}
Model specific parameters:
{\scriptsize
\begin{center}
\begin{tabular}{l|l|c|l}
\bf fomodel/bwa.ini	& \bf in program & \bf symbol	& \bf description\\
\hline
Tau			& mTau		& $\tau$		& proper time\\
VelT			& mVt		& $v_T$			& transverse velocity of the fluid element\\
ParA			& mA		& $A$			& the $A$ parameter in Eq. (\ref{model-a2}) \\
Delay			& mDelay	& $\lambda$		& emission delay\\
\end{tabular}
\end{center}
}
If the {\tt events.ini} file has the {\bf FreezeOutModel} option set to: {\tt BWAVLinear}, {\tt BWAVLinearDelay}, or {\tt BWAVLinearFormation}, the value of the fluid transverse velocity has a linear form
\begin{equation}
  \tilde v_T = \frac{\rho/\rho_{\rm max}}{v_T + \rho/\rho_{\rm max}}.
\end{equation}
Additionally, for the models {\tt BWAVTDelay} and {\tt BWAVLinearDelay} the particle's creation time is shifted randomly according to the exponential distribution
\begin{equation}
  f(t) = e^{-\frac{t-t_0}{\lambda}},
\end{equation}
where $t_0$ is the original time of the particle's creation.\\

The model {\tt BWAVLinearFormation} incorporates the particle's formation time and shifts the position of the particle,
\begin{eqnarray}
  \tilde x = x + \frac{p_x}{E}\, t, \quad \tilde y = y + \frac{p_y}{E}\, t,
\quad \tilde z = z + \frac{p_z}{E}\, t.  \nonumber
\end{eqnarray}
%
%
\section{Freeze-out models based on 2+1 and 3+1 hydrodynamic calculations}
Common parameters in hydro-based models:
{\scriptsize
\begin{center}
\begin{tabular}{l|l|c|l}
\bf lhyquid*.ini	& \bf in program & \bf symbol	& \bf description\\
\hline
RapidityRange		& mRapRange	& $Y^{\rm rng}$	& rapidity range\\
\hline
\multicolumn{3}{c}{}\\
\bf *.xml file	& 		& 			& \\
$<$PARAMETER$>$	&		&			& \\
\hline
Tau\_i		& mTauI		& $\tau_i$		& initial proper time\\
Temperature	& mTemp		& $T$			& freeze-out temperature\\
Mu\_B		& mMuB		& $\mu_B$		& baryon chemical potential\\
Mu\_I		& mMuI		& $\mu_{I_3}$		& isospin chemical potential\\
Mu\_S		& mMuS		& $\mu_S$		& strange chemical potential\\
Mu\_C		& mMuC		& $\mu_C$		& charm chemical potential\\
device			& mDeviceName		& 			& accelerator's name, e.g., RHIC, LHC \\
colliding\_system	& mCollidingSystem	& 			& colliding nuclei, e.g., Au, Pb\\
colliding\_energy	& mCollidingEnergy	& $\sqrt{s_{NN}}$ 	& center-of-mass energy\\
centrality\_min		& mCentralityMin	& $c_{min}$		& centrality lower limit\\
centrality\_max		& mCentralityMax	& $c_{max}$		& centrality higher limit\\
impact\_parameter	& mImpactParameter	& $b$			& impact parameter\\
temperature\_at\_center	& mTempI		& $T_i$			& central initial temperature\\
\hline
\multicolumn{3}{c}{}\\
$<$VECTOR3D$>$	&		&			& \\
\hline
Distance	& mDistance	& $d(\zeta,\phi,\theta)$	& distance to the hypersurface element\\
DistanceDZeta	& mDistanceDZeta	& $ \frac{\partial d}{\partial\zeta}$	& $\zeta$-derivative of $d$\\
DistanceDPhi	& mDistanceDPhi	& $\frac{\partial d}{\partial\phi}$	& $\phi$-derivative of $d$.\\
\end{tabular}
\end{center}
}
Phase-space integration ranges:
{\scriptsize
\begin{center}
\begin{tabular}{rclrcl|rcl}
  $Y_s$		& $\in$ & $\left[ - \frac{ Y_s^{\rm rng}}{2}, \frac{ Y_s^{\rm rng}}{2} \right]$ \mbox{or} &
  $\theta$	& $\in$ & $\left[ \theta_{\rm min}, \theta_{\rm max} \right]$ &
  $Y$		& $\in$ & $\left[ - \frac{ Y^{\rm rng}}{2}, \frac{ Y^{\rm rng}}{2} \right]$ \\
  &&&$\zeta$	& $\in$ & $\left[ \zeta_{\rm min}, \zeta_{\rm max} \right]$ &
  $p_T$		& $\in$ & $\left[ 0 , \infty \right)$ \\
  &&&$\phi$	& $\in$ & $\left[ \phi_{\rm min}, \phi_{\rm max} \right]$ &
  $\phi_p$	& $\in$ & $\left[ 0, 2 \pi \right]$.
\end{tabular}
\end{center}
}
Note that in the {\bf Lhyquid3D} case the space-time rapidity $Y_s$ is parametrized by the $\theta$ angle, $Y_s = (d/\Lambda)\cos\theta $.

\subsection{Lhyquid 3D}
Model specific parameters:
{\scriptsize
\begin{center}
\begin{tabular}{l|l|c|l}
\bf *.xml file	& \bf in program & \bf symbol	& \bf description\\
$<$PARAMETER$>$	&		&			& \\
\hline
Lambda		& mLambda	& $\Lambda$		& conversion constant in $Y_s = (d/\Lambda)\cos\theta $ \\
\hline
\multicolumn{3}{c}{}\\
$<$VECTOR3D$>$	&		&			& \\
\hline
DistanceDTheta	& mDistanceDTheta	& $\frac{\partial d}{\partial\theta}$	& $\theta$-derivative of $d$\\
FluidUx		& mFluidUx	& $u_x(\zeta,\phi,\theta)$	& fluid four-velocity $x$-component\\
FluidUy		& mFluidUy	& $u_y(\zeta,\phi,\theta)$	& fluid four-velocity $y$-component\\
FluidRapidity	& mFluidRapidity	& $Y_f(\zeta,\phi,\theta)$	& fluid rapidity.\\
\end{tabular}
\end{center}
}

\subsection{Lhyquid 2D boost-invariant}
Model specific parameters:
{\scriptsize
\begin{center}
\begin{tabular}{l|l|c|l}
\bf lhyquid2dbi.ini	& \bf in program & \bf symbol	& \bf description\\
\hline
RapSRange		& mRapSRange	& $Y_s^{\rm rng}$	& spatial-rapidity range\\
\hline
\multicolumn{3}{c}{}\\
\bf *.xml file	& 		& 			& \\
$<$VECTOR3D$>$	&		&			& \\
\hline
FluidVt		& mFluidVt	& $v_T(\zeta,\phi)$	& fluid transverse velocity\\
FluidPhi	& mFluidPhi	& $\phi_f(\zeta,\phi)$	& fluid velocity azimuthal angle.\\
\end{tabular}
\end{center}
}


\begin{thebibliography}{100}

\bibitem{Kisiel:2005hn}
A. Kisiel et~al.,
\newblock Comput. Phys. Commun. 174 (2006) 669, nucl-th/0504047.

\bibitem{Broniowski:2005ae}
W. Broniowski et~al.,
\newblock Phys. Lett. B635 (2006) 290, nucl-th/0510033.

\bibitem{Pratt:2005bt}
S. Pratt and D. Schindel,
\newblock AIP Conf. Proc. 828 (2006) 430, nucl-th/0511010.

\bibitem{Andronic:2005yp}
A. Andronic, P. Braun-Munzinger and J. Stachel,
\newblock Nucl. Phys. A772 (2006) 167, nucl-th/0511071.

\bibitem{Kisiel:2006is}
A. Kisiel, W. Florkowski and W. Broniowski,
\newblock Phys. Rev. C73 (2006) 064902, nucl-th/0602039.

\bibitem{Chojnacki:2006tv}
M. Chojnacki and W. Florkowski,
\newblock Phys. Rev. C74 (2006) 034905, nucl-th/0603065.

\bibitem{Florkowski:2006mb}
W. Florkowski et~al.,
\newblock Acta Phys. Polon. B37 (2006) 3381, nucl-th/0609054.

\bibitem{Biedron:2006vf}
B. Biedron and W. Broniowski,
\newblock Phys. Rev. C75 (2007) 054905, nucl-th/0610083.

\bibitem{Utyuzh:2006nw}
O. Utyuzh, G. Wilk and Z. Wlodarczyk,
\newblock Braz. J. Phys. 37 (2007) 708, hep-ph/0610408.

\bibitem{Begun:2006uu}
V.V. Begun et~al.,
\newblock Phys. Rev. C76 (2007) 024902, nucl-th/0611075.

\bibitem{Utyuzh:2007ct}
O. Utyuzh, G. Wilk and Z. Wlodarczyk,
\newblock Phys. Rev. D75 (2007) 074030, hep-ph/0702073.

\bibitem{Brown:2007raa}
D.A. Brown et~al.,
\newblock Phys. Rev. C76 (2007) 044906, 0705.1337.

\bibitem{Baeuchle:2007et}
B. Baeuchle,
\newblock (2007), 0706.4252.

\bibitem{Vertesi:2007ki}
PHENIX, R. Vertesi,
\newblock (2007), 0706.4409.

\bibitem{Danielewicz:2007jn}
P. Danielewicz,
\newblock (2007), 0707.0377.

\bibitem{Begun:2007du}
V.V. Begun,
\newblock Phys. Atom. Nucl. 71 (2008) 1813, 0711.2912.

\bibitem{Tomasik:2007gs}
B. Tomasik et~al.,
\newblock Acta Phys. Polon. Supp. 1 (2008) 513, 0712.0563.

\bibitem{Chojnacki:2007rq}
M. Chojnacki et~al.,
\newblock Phys. Rev. C78 (2008) 014905, 0712.0947.

\bibitem{Bozek:2007qt}
P. Bozek,
\newblock Phys. Rev. C77 (2008) 034911, 0712.3498.

\bibitem{Afanasiev:2007kk}
PHENIX, S. Afanasiev et~al.,
\newblock Phys. Rev. Lett. 100 (2008) 232301, 0712.4372.

\bibitem{Akiba:2008zz}
Y. Akiba,
\newblock Prog. Theor. Phys. Suppl. 174 (2008) 88.

\bibitem{Broniowski:2008vp}
W. Broniowski et~al.,
\newblock Phys. Rev. Lett. 101 (2008) 022301, 0801.4361.

\bibitem{Florkowski:2008cs}
W. Florkowski et~al.,
\newblock Acta Phys. Polon. B39 (2008) 1555, 0804.0974.

\bibitem{Lacey:2008kd}
PHENIX, R.A. Lacey,
\newblock J. Phys. G35 (2008) 104139, 0805.1352.

\bibitem{Chung:2008fu}
P. Chung and P. Danielewicz,
\newblock Phys. Atom. Nucl. 71 (2008) 1552, 0807.4892.

\bibitem{Kisiel:2008ws}
A. Kisiel et~al.,
\newblock (2008), 0808.3363.

\bibitem{:2008fq}
NA49, C. Alt et~al.,
\newblock (2008), 0809.1445.

\bibitem{Tomasik:2008fq}
B. Tomasik,
\newblock Comput. Phys. Commun. 180 (2009) 1642, 0806.4770.

\bibitem{Bozek:2008zw}
P. Bozek,
\newblock Phys. Rev. C79 (2009) 054901, 0811.1918.

\bibitem{:2008gk}
PHOBOS, B. Alver et~al.,
\newblock (2008), 0812.1172.

\bibitem{Broniowski:2008qk}
W. Broniowski et~al.,
\newblock Phys. Rev. C80 (2009) 034902, 0812.3393.

\bibitem{Broniowski:2008ee}
W. Broniowski et~al.,
\newblock Acta Phys. Polon. B40 (2009) 979, 0812.4935.

\bibitem{Florkowski:2009vr}
W. Florkowski et~al.,
\newblock Acta Phys. Polon. B40 (2009) 1093, 0901.1251.

\bibitem{Chajecki:2009zg}
Z. Chajecki,
\newblock Acta Phys. Polon. B40 (2009) 1119, 0901.4078.

\bibitem{Bozek:2009ty}
P. Bozek and I. Wyskiel,
\newblock Phys. Rev. C79 (2009) 044916, 0902.4121.

\bibitem{Luo:2009sx}
X. Luo et~al.,
\newblock Phys. Lett. B673 (2009) 268, 0903.0024.

\bibitem{Bozek:2009zz}
P. Bozek and I. Wyskiel,
\newblock Phys. Rev. C79 (2009) 044916.

\bibitem{NoronhaHostler:2009tz}
J. Noronha-Hostler et~al.,
\newblock (2009), 0906.3960.

\bibitem{Broniowski:2009fm}
W. Broniowski, M. Chojnacki and L. Obara,
\newblock Phys. Rev. C80 (2009) 051902, 0907.3216.

\bibitem{Florkowski:2009rt}
W. Florkowski et~al.,
\newblock Nucl. Phys. A830 (2009) 821c, 0907.3592.

\bibitem{Bozek:2009mz}
P. Bozek and I. Wyskiel,
\newblock (2009), 0909.2354.

\bibitem{Hauer:2009nu}
M. Hauer and S. Wheaton,
\newblock Phys. Rev. C80 (2009) 054915, 0909.2431.

\bibitem{Ryblewski:2009hm}
R. Ryblewski and W. Florkowski,
\newblock (2009), 0910.0985.

\bibitem{Broniowski:2009te}
W. Broniowski et~al.,
\newblock (2009), 0910.3585.

\bibitem{Videbaek:2009zy}
BRAHMS, F. Videbaek,
\newblock Nucl. Phys. A830 (2009) 43c, 0907.4742.

\bibitem{Bozek:2009dt}
P. Bozek,
\newblock (2009), 0911.2392.

\bibitem{Kisiel:2009iw}
A. Kisiel and D.A. Brown,
\newblock (2009), 0901.3527.

\bibitem{Kisiel:2009eh}
A. Kisiel,
\newblock (2009), 0909.5349.

\bibitem{Kisiel:2009kn}
A. Kisiel and T.J. Humanic,
\newblock (2009), 0908.3830.

\bibitem{Florkowski:2009wb}
W. Florkowski and R. Ryblewski,
\newblock (2009), 0912.3451.

\bibitem{Heinz:2009xj}
U.W. Heinz,
\newblock (2009), 0901.4355.

\bibitem{Bozek:2009pu}
P. Bozek and I. Wyskiel,
\newblock (2009), 0903.3129.

\bibitem{Lokhtin:2009hs}
I.P. Lokhtin et~al.,
\newblock (2009), 0910.5129.

\bibitem{NoronhaHostler:2010yc}
J. Noronha-Hostler, J. Noronha and C. Greiner,
\newblock J. Phys. G37 (2010) 094062, 1001.2610.

\bibitem{Luo:2010by}
X.F. Luo et~al.,
\newblock J. Phys. G37 (2010) 094061, 1001.2847.

\bibitem{Bozek:2010bi}
P. Bozek and I. Wyskiel,
\newblock Phys. Rev. C81 (2010) 054902, 1002.4999.

\bibitem{Aggarwal:2010wy}
STAR, M.M. Aggarwal et~al.,
\newblock Phys. Rev. Lett. 105 (2010) 022302, 1004.4959.

\bibitem{Hauer:2010sv}
M. Hauer,
\newblock (2010), 1008.1990.

\bibitem{NoronhaHostler:2010hy}
J. Noronha-Hostler and C. Greiner,
\newblock (2010), 1008.5075.

\bibitem{Bozek:2010aj}
P. Bozek and I. Wyskiel-Piekarska,
\newblock (2010), 1009.0701.

\bibitem{Bilandzic:2010jr}
A. Bilandzic, R. Snellings and S. Voloshin,
\newblock (2010), 1010.0233.

\bibitem{Lizhu:2010zh}
C. Lizhu et~al.,
\newblock (2010), 1011.0712.

\bibitem{Bozek:2010vz}
P. Bozek, W. Broniowski and J. Moreira,
\newblock (2010), 1011.3354.

\bibitem{root}
R. Brun and F. Rademakers,
\newblock Nucl. Instrum. Meth. A389 (1997) 81,
\newblock (http://root.cern.ch).

\bibitem{Florkowski:2001fp}
W. Florkowski, W. Broniowski and M. Michalec,
\newblock Acta Phys. Polon. B33 (2002) 761, nucl-th/0106009.

\bibitem{Broniowski:2001we}
W. Broniowski and W. Florkowski,
\newblock Phys. Rev. Lett. 87 (2001) 272302, nucl-th/0106050.

\bibitem{Danielewicz:1992mi}
P. Danielewicz and Q.b. Pan,
\newblock MSUCL-848.

\bibitem{Schnedermann:1993ws}
E. Schnedermann, J. Sollfrank and U.W. Heinz,
\newblock Phys. Rev. C48 (1993) 2462, nucl-th/9307020.

\bibitem{Retiere:2003kf}
F. Retiere and M.A. Lisa,
\newblock Phys. Rev. C70 (2004) 044907, nucl-th/0312024.

\bibitem{Teaney:2001av}
D. Teaney, J. Lauret and E.V. Shuryak,
\newblock (2001), nucl-th/0110037.

\bibitem{Hirano:2002ds}
T. Hirano and K. Tsuda,
\newblock Phys. Rev. C66 (2002) 054905, nucl-th/0205043.

\bibitem{Kolb:2003dz}
P.F. Kolb and U. Heinz,
\newblock (2003), in Quark-Gluon Plasma 3, edited by R.C. Hwa and X.-N. Wang
  (World Scientific, Singapore, 2004), p. 634, nucl-th/0305084.

\bibitem{Huovinen:2003fa}
P. Huovinen,
\newblock (2003), in Quark-Gluon Plasma 3, edited by R.C. Hwa and X.-N. Wang
  (World Scientific, Singapore, 2004), p. 600, nucl-th/0305064.

\bibitem{Shuryak:2004cy}
E.V. Shuryak,
\newblock Nucl. Phys. A750 (2005) 64, hep-ph/0405066.

\bibitem{Eskola:2005ue}
K.J. Eskola et~al.,
\newblock Phys. Rev. C72 (2005) 044904, hep-ph/0506049.

\bibitem{Hama:2005dz}
Y. Hama et~al.,
\newblock Nucl. Phys. A774 (2006) 169, hep-ph/0510096.

\bibitem{Hirano:2005xf}
T. Hirano et~al.,
\newblock Phys. Lett. B636 (2006) 299, nucl-th/0511046.

\bibitem{Huovinen:2006jp}
P. Huovinen and P.V. Ruuskanen,
\newblock Ann. Rev. Nucl. Part. Sci. 56 (2006) 163, nucl-th/0605008.

\bibitem{Hirano:2007xd}
T. Hirano et~al.,
\newblock J. Phys. G34 (2007) S879, nucl-th/0701075.

\bibitem{Nonaka:2006yn}
C. Nonaka and S.A. Bass,
\newblock Phys. Rev. C75 (2007) 014902, nucl-th/0607018.

\bibitem{Huovinen:2009yb}
P. Huovinen and P. Petreczky,
\newblock (2009), 0912.2541.

\bibitem{Torrieri:2004zz}
G. Torrieri et~al.,
\newblock Comput. Phys. Commun. 167 (2005) 229, nucl-th/0404083.

\bibitem{Torrieri:2006xi}
G. Torrieri et~al.,
\newblock Comput. Phys. Commun. 175 (2006) 635, nucl-th/0603026.

\bibitem{Wheaton:2004qb}
S. Wheaton and J. Cleymans,
\newblock Comput. Phys. Commun. 180 (2009) 84, hep-ph/0407174.

\bibitem{Lokhtin:2008xi}
I.P. Lokhtin et~al.,
\newblock Comput. Phys. Commun. 180 (2009) 779, 0809.2708.

\bibitem{Lokhtin:2009be}
I.P. Lokhtin et~al.,
\newblock (2009), 0903.0525.

\bibitem{Amelin:2006qe}
N.S. Amelin et~al.,
\newblock Phys. Rev. C74 (2006) 064901, nucl-th/0608057.

\bibitem{Amelin:2007ic}
N.S. Amelin et~al.,
\newblock Phys. Rev. C77 (2008) 014903, 0711.0835.

\bibitem{Koch:1985hk}
P. Koch and J. Rafelski,
\newblock South Afr. J. Phys. 9 (1986) 8.

\bibitem{Cleymans:1992zc}
J. Cleymans and H. Satz,
\newblock Z. Phys. C57 (1993) 135, hep-ph/9207204.

\bibitem{Sollfrank:1993wn}
J. Sollfrank et~al.,
\newblock Z. Phys. C61 (1994) 659.

\bibitem{Braun-Munzinger:1994xr}
P. Braun-Munzinger et~al.,
\newblock Phys. Lett. B344 (1995) 43, nucl-th/9410026.

\bibitem{Braun-Munzinger:1995bp}
P. Braun-Munzinger et~al.,
\newblock Phys. Lett. B365 (1996) 1, nucl-th/9508020.

\bibitem{Csorgo:1995bi}
T. Csorgo and B. Lorstad,
\newblock Phys. Rev. C54 (1996) 1390, hep-ph/9509213.

\bibitem{Cleymans:1996cd}
J. Cleymans et~al.,
\newblock Z. Phys. C74 (1997) 319, nucl-th/9603004.

\bibitem{Rafelski:1996hf}
J. Rafelski, J. Letessier and A. Tounsi,
\newblock Acta Phys. Polon. B27 (1996) 1037, nucl-th/0209080.

\bibitem{Rafelski:1997ab}
J. Rafelski, J. Letessier and A. Tounsi,
\newblock Acta Phys. Polon. B28 (1997) 2841, hep-ph/9710340.

\bibitem{Becattini:1997uf}
F. Becattini,
\newblock J. Phys. G23 (1997) 1933, hep-ph/9708248.

\bibitem{Yen:1998pa}
G.D. Yen and M.I. Gorenstein,
\newblock Phys. Rev. C59 (1999) 2788, nucl-th/9808012.

\bibitem{Cleymans:1998fq}
J. Cleymans and K. Redlich,
\newblock Phys. Rev. Lett. 81 (1998) 5284, nucl-th/9808030.

\bibitem{Gazdzicki:1998vd}
M. Gazdzicki and M.I. Gorenstein,
\newblock Acta Phys. Polon. B30 (1999) 2705, hep-ph/9803462.

\bibitem{Gazdzicki:1999ej}
M. Gazdzicki,
\newblock Nucl. Phys. A681 (2001) 153, hep-ph/9910363.

\bibitem{Braun-Munzinger:1999qy}
P. Braun-Munzinger, I. Heppe and J. Stachel,
\newblock Phys. Lett. B465 (1999) 15, nucl-th/9903010.

\bibitem{Cleymans:1999st}
J. Cleymans and K. Redlich,
\newblock Phys. Rev. C60 (1999) 054908, nucl-th/9903063.

\bibitem{Becattini:2000jw}
F. Becattini et~al.,
\newblock Phys. Rev. C64 (2001) 024901, hep-ph/0002267.

\bibitem{Braun-Munzinger:2001ip}
P. Braun-Munzinger et~al.,
\newblock Phys. Lett. B518 (2001) 41, hep-ph/0105229.

\bibitem{Broniowski:2002nf}
W. Broniowski, A. ~ and W. Florkowski,
\newblock Acta Phys. Polon. B33 (2002) 4235, hep-ph/0209286.

\bibitem{Florkowski:2004em}
W. Florkowski, W. Broniowski and P. Bozek,
\newblock J. Phys. G30 (2004) S1321, nucl-th/0403038.

\bibitem{Bozek:2003qi}
P. Bozek, W. Broniowski and W. Florkowski,
\newblock Acta Phys. Hung. A22 (2005) 149, nucl-th/0310062.

\bibitem{Broniowski:2002wp}
W. Broniowski, A. Baran and W. Florkowski,
\newblock AIP Conf. Proc. 660 (2003) 185, nucl-th/0212053.

\bibitem{Florkowski:2004du}
W. Florkowski, W. Broniowski and A. Baran,
\newblock (2004), nucl-th/0412077.

\bibitem{Amsler:2008zzb}
Particle Data Group, C. Amsler et~al.,
\newblock Phys. Lett. B667 (2008) 1.

\bibitem{Hagedorn:1965st}
R. Hagedorn,
\newblock Nuovo Cim. Suppl. 3 (1965) 147.

\bibitem{Hagedorn:1968ua}
R. Hagedorn and J. Ranft,
\newblock Nuovo Cim. Suppl. 6 (1968) 169.

\bibitem{Hagedorn:1994sc}
R. Hagedorn,
\newblock Invited talk at NATO Advanced Study Workshop on Hot Hadronic Matter:
  Theory and Experiment, Divonne-les-Bains, France, 27 Jun - 1 Jul 1994.

\bibitem{Broniowski:2000bj}
W. Broniowski and W. Florkowski,
\newblock Phys. Lett. B490 (2000) 223, hep-ph/0004104.

\bibitem{Broniowski:2000hd}
W. Broniowski,
\newblock (2000), hep-ph/0008112.

\bibitem{Broniowski:2004yh}
W. Broniowski, W. Florkowski and L.Y. Glozman,
\newblock Phys. Rev. D70 (2004) 117503, hep-ph/0407290.

\bibitem{Cooper:1974mv}
F. Cooper and G. Frye,
\newblock Phys. Rev. D10 (1974) 186.

\bibitem{Misner:1974qy}
C.W. Misner, K.S. Thorne and J.A. Wheeler,
\newblock {\rm (W.H. Freeman, San Francisco, 1973) 1279 p} .

\bibitem{Siemens:1978pb}
P.J. Siemens and J.O. Rasmussen,
\newblock Phys. Rev. Lett. 42 (1979) 880.

\bibitem{doxygenweb}
~{\tt http://www.stack.nl/\~{}dimitri/doxygen/}.

\bibitem{torqueweb}
~{\tt http://www.clusterresources.com/products/torque-resource-manager.php}.

\bibitem{Bertsch:1988db}
G. Bertsch, M. Gong and M. Tohyama,
\newblock Phys. Rev. C37 (1988) 1896.

\bibitem{Pratt:1986cc}
S. Pratt,
\newblock Phys. Rev. D33 (1986) 1314.

\bibitem{Bugaev:1996zq}
K.A. Bugaev,
\newblock Nucl. Phys. A606 (1996) 559, nucl-th/9906047.

\bibitem{Bugaev:1999wz}
K.A. Bugaev and M.I. Gorenstein,
\newblock (1999), nucl-th/9903072.

\bibitem{Bugaev:2002ch}
K.A. Bugaev,
\newblock Phys. Rev. Lett. 90 (2003) 252301, nucl-th/0210087.

\bibitem{Colangelo:2001df}
G. Colangelo, J. Gasser and H. Leutwyler,
\newblock Nucl. Phys. B603 (2001) 125, hep-ph/0103088.

\bibitem{Aamodt:2010pa}
The ALICE, K. Aamodt et~al.,
\newblock (2010), 1011.3914.

\bibitem{Bozek:2010er}
P. Bozek,
\newblock (2010), 1012.5927.

\bibitem{Bozek:2011wa}
P. Bozek,
\newblock (2011), 1101.1791.

\bibitem{Letessier:1998sz}
J. Letessier and J. Rafelski,
\newblock Phys. Rev. C59 (1999) 947, hep-ph/9806386.

\bibitem{Muronga:2001zk}
A. Muronga,
\newblock Phys. Rev. Lett. 88 (2002) 062302, nucl-th/0104064.

\bibitem{Teaney:2003kp}
D. Teaney,
\newblock Phys. Rev. C68 (2003) 034913, nucl-th/0301099.

\bibitem{Baier:2006um}
R. Baier, P. Romatschke and U.A. Wiedemann,
\newblock Phys. Rev. C73 (2006) 064903, hep-ph/0602249.

\bibitem{Baier:2006gy}
R. Baier and P. Romatschke,
\newblock Eur. Phys. J. C51 (2007) 677, nucl-th/0610108.

\bibitem{Romatschke:2007mq}
P. Romatschke and U. Romatschke,
\newblock Phys. Rev. Lett. 99 (2007) 172301, 0706.1522.

\bibitem{Chaudhuri:2006jd}
A.K. Chaudhuri,
\newblock Phys. Rev. C74 (2006) 044904, nucl-th/0604014.

\bibitem{Heinz:2005bw}
U.W. Heinz, H. Song and A.K. Chaudhuri,
\newblock Phys. Rev. C73 (2006) 034904, nucl-th/0510014.

\bibitem{Bozek:2007di}
P. Bozek,
\newblock Acta Phys. Polon. B39 (2008) 1375, 0711.2889.

\bibitem{Song:2007ux}
H. Song and U.W. Heinz,
\newblock Phys. Rev. C77 (2008) 064901, 0712.3715.

\bibitem{Bozek:2008bh}
P. Bozek,
\newblock Acta Phys. Polon. B39 (2008) 1539, 0803.4447.

\bibitem{Song:2008hj}
H. Song and U.W. Heinz,
\newblock (2008), 0812.4274.

\bibitem{Luzum:2008cw}
M. Luzum and P. Romatschke,
\newblock Phys. Rev. C78 (2008) 034915, 0804.4015.

\bibitem{Bozek:2009ut}
P. Bozek,
\newblock Acta Phys. Polon. B40 (2009) 987, 0901.2272.

\bibitem{Bozek:2009dw}
P. Bozek,
\newblock (2009), 0911.2397.

\bibitem{Romatschke:2009im}
P. Romatschke,
\newblock (2009), 0902.3663.

\bibitem{Brown:1997ku}
D.A. Brown and P. Danielewicz,
\newblock Phys. Lett. B398 (1997) 252, nucl-th/9701010.

\bibitem{Hagiwara:2002fs}
Particle Data Group, K. Hagiwara et~al.,
\newblock Phys. Rev. D66 (2002) 010001.

\end{thebibliography}
\end{document}